\documentclass[12pt]{article}
\usepackage{latexsym,amssymb,amsmath}
\begin{document}
\baselineskip=20pt

\newcommand{\la}{\langle}
\newcommand{\ra}{\rangle}
\newcommand{\psp}{\vspace{0.4cm}}
\newcommand{\pse}{\vspace{0.2cm}}
\newcommand{\ptl}{\partial}
\newcommand{\dlt}{\delta}
\newcommand{\sgm}{\sigma}
\newcommand{\al}{\alpha}
\newcommand{\be}{\beta}
\newcommand{\G}{\Gamma}
\newcommand{\gm}{\gamma}
\newcommand{\vs}{\varsigma}
\newcommand{\Lmd}{\Lambda}
\newcommand{\lmd}{\lambda}
\newcommand{\td}{\tilde}
\newcommand{\vf}{\varphi}
\newcommand{\yt}{Y^{\nu}}
\newcommand{\wt}{\mbox{wt}\:}
\newcommand{\rd}{\mbox{Res}}
\newcommand{\ad}{\mbox{ad}}
\newcommand{\stl}{\stackrel}
\newcommand{\ol}{\overline}
\newcommand{\ul}{\underline}
\newcommand{\es}{\epsilon}
\newcommand{\dmd}{\diamond}
\newcommand{\clt}{\clubsuit}
\newcommand{\vt}{\vartheta}
\newcommand{\ves}{\varepsilon}
\newcommand{\dg}{\dagger}
\newcommand{\tr}{\mbox{Tr}}
\newcommand{\ga}{{\cal G}({\cal A})}
\newcommand{\hga}{\hat{\cal G}({\cal A})}
\newcommand{\Edo}{\mbox{End}\:}
\newcommand{\for}{\mbox{for}}
\newcommand{\kn}{\mbox{ker}}
\newcommand{\Dlt}{\Delta}
\newcommand{\rad}{\mbox{Rad}}
\newcommand{\rta}{\rightarrow}
\newcommand{\mbb}{\mathbb}
\newcommand{\lra}{\Longrightarrow}

\begin{center}{\Large \bf New Algebraic Approaches to
 Classical }\end{center}
\begin{center}{\Large \bf   Boundary Layer Problems}
\end{center}
\vspace{0.2cm}

\begin{center}{\large Xiaoping Xu}\end{center}
\begin{center}{Institute of Mathematics, Academy of Mathematics \& System Sciences}\end{center}
\begin{center}{Chinese Academy of Sciences, Beijing 100080, P.R. China}
\footnote{Research supported
 by China NSF 10431040}\end{center}

\vspace{0.6cm}

 \begin{center}{\Large\bf Abstract}\end{center}

\vspace{1cm} {\small In this paper, we use various ansatzes with
undetermined functions and the technique of moving frame to find
solutions with parameter functions modulo the Lie point symmetries
for the classical non-steady boundary layer problems. These
parameter functions enable one to find the solutions of some
related practical models and boundary value problems.}

\vspace{0.8cm}

\section{Introduction}

In 1904, Prandtl observed that in the flow of slightly viscous
fluid past bodies, the frictional effects are confined to a thin
layer of fluid adjacent to the surface of the body. Moreover, he
showed that the motion of a small amount of fluid in this boundary
layer decides such important matters as the frictional drag, heat
transfer, and transfer of momentum between the body and the fluid.
 The two-dimensional classical non-steady boundary layer
equations
$$u_t+uu_x+vu_y+p_x=u_{yy},\eqno(1.1)$$
$$p_y=0,\qquad u_x+v_y=0\eqno(1.2)$$
are used to describe the motion of a flat plate with the incident
flow parallel to the plate and directed to along the $x$-axis in
the Cartesian coordinates $(x,y)$, where $u$ and $v$ are the
longitudinal and the transverse components of the velocity, and
$p$ is the pressure.  Lie point symmetries of the above equations
were obtained by Ovsiannikov [O]. The three-dimensional classical
non-steady boundary layer equations are:
$$u_t+uu_x+vu_y+wu_z=-\frac{1}{\rho}p_x+\nu u_{yy},\eqno(1.3)$$
$$w_t+uw_x+vw_y+ww_z=-\frac{1}{\rho}p_z+\nu w_{yy},\eqno(1.4)$$
$$p_y=0,\qquad u_x+v_y+w_z=0,\eqno(1.5)$$
where $\rho$ is the density constant and $\nu$ is the coefficient
constant of the kinematic viscosity. The Lie point symmetries of
the equations (1.3)-(1.5) were obtained by Garaev [G] and by
Vereschagina [V]. Lie group method is one of most important ways
of solving nonlinear partial differential equations. However, the
method only enables one to obtain certain special solutions. It is
desirable to find more effective ways of solving nonlinear partial
differential equations.

The aim of this paper is to develop more powerful methods than the
Lie group method of solving the above classical non-steady
boundary layer problems. Based our earlier work [X] on
function-parameter exact solutions of the equation of
nonstationary transonic gas flow discovered by Lin, Reisner and
Tsien [LRT], and its three-dimensional generalization, we find
ansatzes with various undetermined functions and obtained
solutions with parameter functions modulo the Lie point symmetries
for the above boundary layer equations. In the three-dimensional
case, we have also used the technique of moving frame. Our results
can be applied to certain related practical models and boundary
value problems by specifying these parameter functions. Below we
give a more detailed introduction.

Lin, Reisner and Tsien [LRT] found the  equation
$$2u_{tx}+u_xu_{xx}-u_{yy}=0\eqno(1.6)$$
for two-dimensional non-steady motion of a slender body in a
compressible fluid, which was later called the ``equation of
nonstationary transonic gas flows" (cf. [M]). Using certain
finite-dimensional stable range of the nonlinear term $u_xu_{xx}$,
we obtain in [X] the following solution of the above equation:
$$u=\frac{x^3}{(\sqrt{3}y+\be)^2} +\frac{\be'x^2}{\sqrt{3}y+\be}
-\frac{2{\be'}'}{3}(\sqrt{3}y+\be)x -\frac{4\be'{\be'}'}{3}y^2
-\frac{2{{\be'}'}'}{27}(\sqrt{3}y+\be)^3,\eqno(1.7)$$ where $\be$
is an arbitrary function of $t$. Since the above solution blows up
on the moving line $\sqrt{3}y+\be=0$, it reflects partial
phenomena of gust. Moreover, we have found a family of solutions
of the three-dimensional generalization of the equation (1.6)
blowing up on a rotating and translating plane
$y\cos\al(t)+z\sin\al(t)=f(t)$, which reflect partial phenomena of
turbulence. Our work [X] suggested some possible approaches to
more general nonlinear partial differential equations in fluid
dynamics, such as the classical non-steady boundary layer problems
we will deal with in this paper.

By the divergence-free condition in (1.2), we write $u$ and $v$ in
the potential form:
$$u=\xi_y,\qquad v=-\xi_x,\eqno(1.8)$$
where $\xi$ is a function of $t,x,y$. It is known that the
solution space of the equations (1.1) and (1.2) is invariant under
the following transformation $T$ (cf. [O]):
$$T(u)=u(t,x,y+\sgm),\qquad
T(v)=v(t,x,y+\sgm)-\sgm_t-\sgm_x u ,\qquad T(p)= p\eqno(1.9)$$ for
any function $\sgm$ of $t,x$. Modulo this transformation, our main
method of solving the equations (1.1) and (1.2) is to assume
$$\xi=f(t,x)+\kappa(t,x)y+g(t,x)H(t,y),\eqno(1.10)$$
where $f,\kappa,g$ are functions to be determined and $H$ is a
more sophisticated function that is either first given or is to be
solved under certain conditions. The term $\kappa(t,x)y$ is used
to make the function $H$ more flexible. With the ansatz (1.10),
the equation (1.1) is reduced to a system of several nonlinear
partial differential equations in $t,x$. While by Lie group
method, one usually reduces it to a single nonlinear partial
differential equation in $t,x$. This in a way shows that our
method is more powerful than the Lie group method. Imposing some
conditions on the above undetermined functions, we are able to
solve the system of partial differential equations. In this way,
we obtain three families of function-parameter exact solutions of
the two-dimensional classical non-steady boundary layer problem
(1.1) and (1.2). For instance, we have the following solutions
(see Theorem 3.2):
$$u=\frac{{\be'}'x}{2\be'}+
\frac{e^{-\be}}{\sqrt{\be'}}
\Im\left(\frac{x}{\sqrt{\be'}}\right)\cos(\sqrt{\be'}\:y),
\eqno(1.11)$$
$$v=-\frac{{\be'}'y}{2\be'}-
\frac{e^{-\be}}{\sqrt{(\be')^3}}
\Im'\left(\frac{x}{\sqrt{\be'}}\right)\sin(\sqrt{\be'}\:y),
\eqno(1.12)$$
$$p=\frac{({\be'}')^2-2\be'{{\be'}'}'}{4(\be')^2}x^2
-\frac{1}{2\be'}e^{-2\be}
\left[\Im\left(\frac{x}{\sqrt{\be'}}\right)\right]^2,\eqno(1.13)$$
where $\be$ is an arbitrary increasing function of $t$ and $\Im$
is any one-variable function. Applying the symmetry
transformations of the equations (1.1) and (1.2) (see next
section) such as the one in (1.9), we can get solutions with more
parameter functions.

The main contents of this paper are devoted to solving the
three-dimensional classical non-steady boundary layer equations
(1.3)-(1.5). Similarly, we write $u,v,w$ in the potential form:
$$u=\xi_y,\qquad w=\eta_y,\qquad v=-(\xi_x+\eta_z)\eqno(1.4)$$
 for some functions $\xi$ and $\eta$ in $t,x,y,z$. The equations
 (1.3)-(1.6) have the similar symmetry transformation as that in
 (1.9). Modulo this transformation, we assume
$$\xi=f(t,x,z)+\kappa(t,x,z)y+g(t,x,z)H(t,\sgm(t,x,z)y),\eqno(1.14)$$
and
$$\eta=\tau(t,x,z)y+\zeta(t,x,z)
\Phi(t,\sgm(t,x,z)y),\eqno(1.15)$$ where
$f,g,\kappa,\sgm,\tau,\zeta$ are three-variable functions to be
determined, and $H(t,\varpi)$ and $\Phi(t,\varpi)$ are more
sophisticated two-variable functions that are either first given
or are to be solved under certain conditions. Again the terms
$\kappa(t,x,z)y$ and $\tau(t,x,z)y$ are used to  make $H$ and
$\Phi$ more flexible. With the ansatz (1.14) and (1.15), the
equations (1.3) and (1.4) are reduced to a system of several more
complicated nonlinear partial differential equations in $t,x,z$.
This again in a way shows that our method is much more powerful
than the Lie group method. One of the main technique of solving
the system of nonlinear partial differential equations in $t,x,z$
is to use time-dependent orthogonal transformations (moving
frames). Imposing some technical conditions on the above
undetermined functions, we are able to solve the system of partial
differential equations. In this way, we obtain ten families of
function-parameter exact solutions of the three-dimensional
classical non-steady boundary layer problem (1.3)-(1.5). For
instance, we have the following solutions: (1) for any function
$\iota$ of $z$, any function $\al$ of $t$ and arbitrary functions
  $\Im,\vf$ in $\hat\varpi=\al-\iota$:
$$u=\frac{{\al'}'}{2\al'}x-\frac{\vf}{\sqrt{\al'}}
\left[\exp\left(-2\nu\Im^2\int\frac{dz}{(\iota')^2}\right)
\right]\cos\left(\frac{\sqrt{\al'}\Im
y}{\iota'}\right),\eqno(1.16)$$ $$ v=
\frac{\al'{\iota'}'y}{(\iota')^2}-\frac{{\al'}'}{2\al'}y,\qquad
w=\frac{\al'}{\iota'},\eqno(6.17)$$
$$p=-\frac{\rho}{2}\left(\frac{2\al'{{\al'}'}'-({\al'}')^2}
{4(\al')^2}x^2+\frac{(\al')^2}
{(\iota')^2}+{\al'}'\int\frac{dz}{\iota'}\right)\eqno(6.18)$$ (see
Theorem 6.1); (2)\hspace{0.2cm} in terms of the moving frame
$$\td\varpi=x\cos\gm+z\sin\gm,\qquad \hat\varpi=z\cos\gm-x\sin\gm,\eqno(1.19)$$
\begin{eqnarray*}\hspace{1cm}u&=&\frac{\be^2\gm'+2\nu(\be'\cos\gm-2\be\gm'\sin\gm-
6\gm')}{4\nu(\be-6\nu\sin\gm)}(\td\varpi+2\hat\varpi\tan\gm)\\
&
&-\left(\frac{1}{6\nu}\be\gm'\tan\gm+\gm'\sec\gm\right)\hat\varpi
-(6\nu x+\be\hat\varpi)y^{-2},\hspace{3.9cm}(1.20)\end{eqnarray*}
$$v=\frac{\be^2\gm'+2\nu(\be'\cos\gm-2\be\gm'\sin\gm-
6\gm')}{4\nu\cos\gm\;(\be-6\nu\sin\gm)}y-
\frac{\be\gm'}{6\nu}y\sec\gm-6\nu y^{-1}, \eqno(1.21)$$
\begin{eqnarray*}\hspace{1cm}w&=&\frac{\be^2\gm'+2\nu(\be'\cos\gm-2\be\gm'\sin\gm-
6\gm')}{4\nu(\be-6\nu\sin\gm)}(\td\varpi\tan\gm-2\hat\varpi)\\
& &+\frac{1}{6\nu}\be\gm'\hat\varpi-\gm'\td\varpi\sec\gm- (6\nu
x+\be\hat\varpi)y^{-2}\tan\gm\hspace{4.5 cm}(1.22)\end{eqnarray*}
 and $p$ is
more complicated and given in (7.26), where $\be$ and $\gm$ are
arbitrary functions of $t$ (see Theorem 7.1).  Applying the
symmetry transformations of the equations (1.3)-(1.5) (see next
section) such as that similar to the one in (1.9), we can get
solutions with more parameter functions. For instance, the above
second solution would yield to a solution singular on any moving
hypersurface, which may be used to study turbulence. In addition,
we present another two anzatzes and obtain another two families of
function-parameter exact solutions of the three-dimensional
problem that are rational functions in  $y$.

Of course, one can view the two-dimensional problem (1.1) and
(1.2) as a special case of the three-dimensional problem
(1.3)-(1.5). Our results on the two-dimensional problem in Section
3 can be directly applied to the three dimensional problem by
assuming that $\xi$ is independent of $z$ and $\eta$ is
independent of $x$. Applying constant orthogonal transformations
(see Section 2), we can obtain more sophisticated solutions of the
three-dimensional problem. So obtained solutions of the
three-dimensional problem will not be given. Thus not all the
solutions of the two-dimensional problem in Section 3 are the
specializations of the solutions of the three-dimensional problem
in later sections. Our another purpose of presenting the solutions
 of the two-dimensional problem in Section 3 is to let the reader
 first understand our techniques in simpler version and then to better
 understand much more sophisticated techniques of solving the
 three-dimensional problem in later sections. Indeed, due to the
 technicality, the author himself keeps referring the results and
 arguments on the two-dimensional problem in Section 3 when he
 solves the three-dimensional problem in later sections.

For convenience, we always assume that all the involved partial
derivatives of related functions always exist and we can change
orders of taking partial derivatives. As we all know that in
general,  it is impossible to find all the solutions of nonlinear
partial differential equations both analytically and
algebraically. In our arguments throughout this paper, we always
search for reasonable sufficient conditions of obtaining exact
solutions. For instance, we treat nonzero functions as ``nonzero
constants" for this purpose because our approaches in this paper
are completely algebraic. Of course, one can used our methods in
this paper to get more solutions, in particular, by considering
the support and discontinuity of the related functions.

The paper is organized as follows. In Section 2, we present the
known Lie point symmetries of the equations (1.1) and (1.2) due to
Ovsiannikov [O], and the Lie point symmetries of the equations
(1.3)-(1.5) due to Garaev [G] and by Vereschagina [V]. In later
sections, we always try to exclude the obvious ingredients in our
solutions induced by these symmetry transformations. In other
words, we search for the solutions modulo these transformations.
We solve the two-dimensional classical non-steady boundary layer
equations (1.1) and (1.2) in Section 3. Starting from Section 4,
we solve the three-dimensional classical non-steady boundary layer
equations (1.3)-(1.5) according to the types of the involved
functions with respect to the variable $y$. First in Section 4, we
find solutions of the equations (1.3)-(1.5) containing a single
exponential function in $y$. In Section 5, we look for solutions
of the equations (1.3)-(1.5) with distinct exponential functions
in $y$. In Section 6, we obtain solutions of the equations
(1.3)-(1.5) with trigonometric and hyperbolic-type functions in
$y$. Finally in Section 7, we present solutions the equations
(1.3)-(1.5) which are rational functions with respect to the
variable $y$.

\section{Lie Point Symmetries}

In this section, we will present all the known Lie point
symmetries of concerned boundary layer problems.

Denote by ${\cal F}_{(i)}(x_1,...,x_n)$ the space of functions in
$x_1,...,x_n$, whose all partial derivatives up $i$th order
exists. Ovsiannikov [O] found the following symmetric group $G_2$
generated by the transformations
$$\{T_{1a},T_{2b},T_{3b},T_{4\al},T_{5\be},T_{6\sgm}\mid
a,b\in\mbb{R},b\neq 0;\al\in {\cal F}_{(2)}(t);\be\in{\cal
F}_{(0)}(t);\sgm\in {\cal F}_{(1)}(t,x)\}\eqno(2.1)$$ with
$$T_{1a}(u)=u(t+at,x,y),\qquad T_{1a}(v)=
v(t+at,x,y),\qquad T_{1a}(p)= p(t+at,x);\eqno(2.2)$$
$$T_{2b}(u)=u(b^2t,b^2x,by),\qquad T_{2b}(v)=b^{-1}
v(b^2t,b^2x,by),\qquad T_{2b}(p)= p(b^2t,b^2x);\eqno(2.3)$$
$$T_{3b}(u)=bu(t,bx,by),\qquad T_{3b}(v)=v(t,bx,by)
,\qquad T_{3b}(p)= b^2p(t,bx);\eqno(2.4)$$
$$T_{4\al}(u)=u(t,x+\al,y)-\al',\;\; T_{4\al}(v)=v(t,x+\al,y)
,\;\; T_{4\al}(p)= p(t,x+\al)+{\al'}'x;\eqno(2.5)$$
$$T_{5\be}(u)=u,\qquad T_{5\be}(v)=v
,\qquad T_{5\be}(p)= p+\be;\eqno(2.6)$$
$$T_{6\sgm}(u)=u(t,x,y+\sgm),\qquad
T_{6\sgm}(v)=v(t,x,y+\sgm)-\sgm_t-\sgm_x u ,\qquad T_{6\sgm}(p)=
p.\eqno(2.7)$$ The elements of $G_2$ transform the solutions of
the two-dimensional classical non-steady boundary layer equations
(1.1) and (1.2) into solutions. For instance, the transformation
$T_{4\al}T_{5\gm}$ transforms a solution with zero pressure into a
solution with the pressure $p={\al'}'x+\gm$ and the transformation
$T_{6\sgm}$ may change a solution to the one with an additional
two-variable parameter function. In this way, we can always
transform our solutions to the ones with nonzero pressure and at
least one two-variable parameter function. For simplicity, we only
write our solutions of (1.1) and (1.2) as simple as possible
modulo the group $G_2$. We will do the same thing for the
solutions of the three dimensional problems (1.3)-(1.5).

The symmetry group $G_3$ of the equations (1.3)-(1.5) are
generated by the transformations
\begin{eqnarray*}\hspace{1cm}& &\{T_{1a},T_{2a},T_{3b},T_{4b},
T_{5\al},T_{6\al},T_{7\be},T_{8\sgm}\mid a,b\in\mbb{R},b\neq 0;\\
& &\al\in {\cal F}_{(2)}(t);\be\in{\cal F}_{(0)}(t); \sgm\in {\cal
F}_{(1)}(t,x,z)\}\hspace{6.2cm}(2.8)\end{eqnarray*} with
$$T_{1a}(\psi)=\psi(t+a,x,z,y),\qquad \psi=u,v,w,p;\eqno(2.9)$$
\begin{eqnarray*}\hspace{1cm}T_{2a}(u)&=&u(t,x\cos a+z\sin a,-x\sin a+z\cos
a,y)\cos a\\ & &+w(t,x\cos a+z\sin a,-x\sin a+z\cos a,y)\sin
a,\hspace{3.1cm}(2.10)\end{eqnarray*}
\begin{eqnarray*}\hspace{1cm}T_{2a}(w)&=&-u(t,x\cos a+z\sin a,-x\sin a+\cos
a,y)\sin a\\ & &+w(t,x\cos a+z\sin a,-x\sin a+z\cos a,y)\cos
a,\hspace{2.9cm}(2.11)\end{eqnarray*}
$$T_{2a}(\psi)=\psi(t,x\cos a+z\sin a,-x\sin a+z\cos
a,y),\qquad \psi=v,p;\eqno(2.12)$$
$$T_{3b}(u)=bu(t,bx,y,bz),\qquad T_{3b}(v)=
v(t,bx,y,bz),\eqno(2.13)$$ $$T_{3b}(w)=bw(t,bx,y,bz),\qquad
T_{3b}(p)= b^2p(t,bx,y,bz);\eqno(2.14)$$
$$T_{4b}(u)=b^{-2}u(b^2t,x,by,z),\qquad T_{4b}(v)=
b^{-1}v(b^2t,x,by,z),\eqno(2.15)$$
$$T_{4b}(w)=b^{-2}w(b^2t,x,by,z),\qquad T_{4b}(p)=
b^{-4}p(b^2t,x,by,z);\eqno(2.16)$$
$$T_{5\al}(u)=u(t,x+\al,y,z)-\al',\qquad
T_{5\al}(v)=v(t,x+\al,y,z),\eqno(2.17)$$
$$T_{5\al}(w)=w(t,x+\al,y,z),\qquad T_{5\al}(p)=
p(t,x+\al,z)+\rho{\al'}'x;\eqno(2.18)$$
$$T_{6\al}(w)=w(t,x,y,z+\al)-\al',\qquad
T_{6\al}(v)=v(t,x,y,z+\al),\eqno(2.19)$$
$$T_{6\al}(u)=u(t,x,y,z+\al),\qquad T_{6\al}(p)=
p(t,x,z+\al)+\rho {\al'}'z;\eqno(2.20)$$
$$T_{7\be}(u)=u,\qquad T_{7\be}(v)=v,\qquad T_{7\be}(w)=w,\qquad T_{7\be}(p)=
p+\be;\eqno(2.21)$$
$$T_{8\sgm}(u)=u(t,x,y+\sgm,z),\qquad
T_{8\sgm}(v)=v(t,x,y+\sgm)-\sgm_t-\sgm_x u -\sgm_z v,\eqno(2.22)$$
$$T_{8\sgm}(u)=w(t,x,y+\sgm,z),\qquad T_{8\sgm}(p)= p.\eqno(2.23)$$

\section{Solutions for the 2-D Problem}

In this section, we will find various function-parameter exact
solutions for the two-dimensional classical non-steady boundary
layer equations (1.1) and (1.2).

Solving the second equation in (1.2),  we get
$$u=\xi_y,\qquad v=-\xi_x\eqno(3.1)$$
for some function $\xi$ in $t,x,y$. Now the equation (1.1) becomes
$$\xi_{yt}+\xi_y\xi_{xy}-\xi_x\xi_{yy}+p_x=\xi_{yyy}.\eqno(3.2)$$
Assume that  $\xi$ is of the form:
$$\xi=f(t,x)+\kappa(t,x)y+g(t,x)H(t,y),\eqno(3.3)$$
where $\kappa(t,x),g(t,x),f(t,x)$ and $H(t,y)$ are two-variable
functions. Then
$$\xi_y=\kappa+g H_y,\qquad \xi_x=f_x+\kappa_xy+g_xH,\qquad
\xi_{yt}=\kappa_t+g_tH_y+gH_{yt},\eqno(3.4)$$
$$\xi_{yx}=\kappa_x+g_xH_y,\qquad
\xi_{yy}=gH_{yy},\qquad\xi_{yyy}=gH_{yyy}.\eqno(3.5)$$ In
particular, the nonlinear term in (3.2) becomes
\begin{eqnarray*}& &\xi_y\xi_{xy}-\xi_x\xi_{yy}=
(\kappa+gH_y)(\kappa_x+g_xH_y)-(f_x+\kappa_xy+g_xH)gH_{yy}\\
&=&\kappa\kappa_x+ (\kappa
g_x+\kappa_xg)H_y-(f_x+\kappa_xy)gH_{yy}+gg_x[H_y^2
-HH_{yy}].\hspace{3.5cm}(3.6)
\end{eqnarray*}

In order to linearize (3.6) with respect to $H$, we divided it
into the following cases.

First we consider $H=e^{\gm y}$.  Then
$$\xi_{yt}=\kappa_t+(\gm g_t+\gm' g)e^{\gm y}+\gm\gm' gye^{\gm
y}.\eqno(3.7)$$ So (3.2) becomes
$$\kappa_t+\kappa\kappa_x+[\gm'g+\gm(g_t+\gm' gy+\kappa g_x+\kappa_xg-\gm f_x-\gm \kappa_xgy-\gm^2
g)]e^{\gm y}+p_x=0,\eqno(3.8)$$ which is implied by the following
system of partial differential equations:
$$\kappa_t+\kappa\kappa_x+p_x=0,\qquad \gm'-\gm \kappa_x=0,\eqno(3.9)$$
$$\gm'g+\gm g_t+\kappa\gm g_x+\gm \kappa_xg-\gm^3 g-\gm ^2gf_x=0.\eqno(3.10)$$
So
$$\kappa=\frac{\gm'}{\gm}x+\gm_0\eqno(3.11)$$
for some function $\gm_0$ of $t$. Moreover, (3.10) implies
$$f_x=\frac{2\gm'}{\gm^2}+\frac{g_t+\gm_0 g_x}{\gm g}
+\frac{\gm'xg_x}{\gm^2g}-\gm.\eqno(3.12)$$
 So
$$\xi=f+\frac{\gm'}{\gm}xy+\gm_0y+ge^{\gm y}.\eqno(3.13)$$ Hence,
$$p=-\frac{{\gm'}'}{2\gm}x^2-\frac{\gm_0'\gm+\gm_0\gm'}{\gm}x
 \eqno(3.14)$$ modulo some
$T_{5\be}$ in (2.6).

Next we consider the case $f=g_x=0$. Replacing $H$ by $gH$, we can
assume $g=1$. Now (3.2) becomes the following linear partial
differential equation:
$$\kappa_t+\kappa\kappa_x+H_{yt}+ \kappa_xH_y-\kappa_xyH_{yy}+p_x =H_{yyy}.
\eqno(3.15)$$ To simplify the problem, we assume
$$\kappa_t+\kappa\kappa_x+p_x=0,\qquad \kappa_x=\frac{{\gm'}'}{2\gm'}
\eqno(3.16)$$ for some nonzero function $\gm'$ of $t$. Then
$$\kappa=\frac{{\gm'}'}{2\gm'}x+\gm_0\eqno(3.17)$$
for some function $\gm_0$ of $t$ and
$$p=\frac{2\gm'{{\gm'}'}'-({\gm'}')^2}{8(\gm')^2}x^2
+\frac{\gm_0{\gm'}'+\gm_0'\gm'}{2\gm'}x\eqno(3.18)$$ modulo the
transformation in (2.6). Now (3.15) becomes
$$H_{yt}+\frac{{\gm'}'}{2\gm'}(H_y-yH_{yy}) = H_{yyy}.\eqno(3.19)$$
Write
$$H_y=\frac{1}{\sqrt{\gm'}}\hat H(t,\sqrt{\gm'}\:y)\eqno(3.20)$$
for some two variable function $\hat H(t,\varpi)$ with
$\varpi=\sqrt{\gm'}\:y$. Then (3.19) turns out to be
$$\hat H_t=\gm'\hat H_{\varpi\varpi}.\eqno(3.21)$$
One can use Fourier expansion to solve the above equation with
given initial condition. In particular, discontinuous solutions
can be found in this way. For the algebraic neatness, we just want
to find globally analytic solution with respect to the spacial
variables $x$ and $y$. So we take the following solution of
(3.21):
$$\hat
H=\sum_{i=1}^md_ie^{(a_i^2-b_i^2)\gm+a_i\varpi}\sin(b_i(2a_i\gm
+\varpi)+c_i),\eqno(3.22)$$ where $a_i,b_i,c_i,d_i$ are real
numbers. Hence, we have
$$H_y=\frac{1}{\sqrt{\gm'}}\sum_{i=1}^md_ie^{(a_i^2-b_i^2)\gm
+a_i\sqrt{\gm'}\:y}\sin(b_i(2a_i\gm
+\sqrt{\gm'}\:y)+c_i).\eqno(3.23)$$

 By (3.1), we have the following result, where the first solution
follows from a direct observation the equations in (1.1) and
(1.2). \psp

{\bf Theorem 3.1}. {\it  We have the following solutions of the
two-dimensional classical non-steady boundary layer problem (1.1)
and (1.2):  (1)
$$u=y,\qquad v=-\kappa_x,\qquad p=\kappa,\eqno(3.24)$$
where $\kappa$ is an arbitrary function of $t,x$; (2)
$$u=\frac{\gm'x}{\gm}+\gm_0+\gm ge^{\gm y},\eqno(3.25)$$
$$v=\gm-\frac{2\gm'}{\gm^2}-\frac{g_t+\gm_0
g_x}{\gm g} -\frac{\gm'xg_x}{\gm^2g}-\frac{\gm'y}{\gm} -g_xe^{\gm
y}.\eqno(3.26)$$ and $p$ is given in (3.14), where $\gm,\gm_0$ are
any nonzero functions of $t$ and $g$ is any nonzero function of
$t,x$; (3)
$$u=\frac{{\gm'}'x}{2\gm'}+\gm_0+\frac{1}{\sqrt{\gm'}}\sum_{i=1}^md_ie^{(a_i^2-b_i^2)\gm
+a_i\sqrt{\gm'}\:y}\sin(b_i(2a_i\gm
+\sqrt{\gm'}\:y)+c_i),\eqno(3.27)$$
$$v=-\frac{{\gm'}'y}{2\gm'}\eqno(3.28)$$  and $p$ is
given in (3.18), where $\gm,\gm_0$ are any functions of $t$, and
$a_i,b_i,c_i,d_i$ are real constants.} \psp

Next we consider the case $f=\ptl_y(H_y^2-H H_{yy})=0$. Again let
$\gm$ be a function of $t$. For $a\in\mbb{R}$, we denote
$$\vt_0=\frac{e^{\gm y}-ae^{-\gm y}}{2},\qquad \vt_1=\sin(\gm
y),\eqno(3.29)$$
$$\hat\vt_0=\frac{e^{\gm y}+ae^{-\gm y}}{2},\qquad \hat\vt_1=\cos(\gm
y).\eqno(3.30)$$ Given $r\in\{0,1\}$.
 Assume
$$H=\vt_r.\eqno(3.31)$$
 Then
$$H_y^2-HH_{yy}=(\dlt_{0,r}a+\dlt_{1,r})\gm^2.
\eqno(3.32)$$ By (3.6),
$$\xi_y\xi_{yx}-\xi_x\xi_{yy}=\kappa\kappa_x+ (\kappa
g_x+\kappa_xg)\gm\hat\vt_r-(-1)^r\gm^2(f_x+\kappa_xy)g\vt_r+(\dlt_{0,r}a+\dlt_{1,r})
\gm^2gg_x.\eqno(3.33)$$ Since
$$\xi_{yt}=\kappa_t+(g_t\gm+g\gm')\hat\vt_r-(-1)^r\gm'\gm g y\vt_r
,\qquad\xi_{yyy}=(-1)^r\gm^3\hat\vt_r,\eqno(3.34)$$
\begin{eqnarray*}\hspace{1.5cm}& &\kappa_t+\kappa\kappa_x+p_x+
+ [\gm'g+\gm(g_t+\kappa g_x+\kappa_xg-(-1)^r\gm^2g)]\hat\vt_r\\
& &-(-1)^r\gm(\gm'+\gm\kappa_x)y g\vt_r+
(\dlt_{0,r}a+\dlt_{1,r})\gm^2gg_x=0,\hspace{3.6cm}(3.35)\end{eqnarray*}
which is implied by the following equations:
$$\kappa_t+\kappa\kappa_x+(\dlt_{0,r}a+\dlt_{1,r})\gm^2gg_x
+p_x=0,\qquad \gm \kappa_x+\gm'=0,\eqno(3.36)$$
$$g\gm'+(g_t+\kappa g_x+\kappa_xg-(-1)^r\gm^2g)\gm=0.\eqno(3.37)$$
For convenience of solving the above equations, we assume
$$\gm=\sqrt{\be'}\lra
\frac{\gm'}{\gm}=\frac{{\be'}'}{2\be'}.\eqno(3.38)$$ So we take
$$\kappa=\frac{{\be'}'x}{2\be'}
,\qquad
g=\frac{e^{(-1)^r\be}}{\be'}\Im\left(\frac{x}{\sqrt{\be'}}\right)
\eqno(3.39)$$ for some one-variable function $\Im$. Moreover,
$$p=\frac{({\be'}')^2-2\be'{{\be'}'}'}{4(\be')^2}x^2
-\frac{\dlt_{0,r}a+\dlt_{1,r}}{2\be'}e^{(-1)^r2\be}
\left[\Im\left(\frac{x}{\sqrt{\be'}}\right)\right]^2\eqno(3.40)$$
modulo the transformation in (2.6) and
$$\xi=\frac{{\be'}'xy}{2\be'}+
\frac{e^{(-1)^r\be}}{\be'}\Im\left(\frac{x}{\sqrt{\be'}}\right)\vt_r.
\eqno(3.41)$$

 According
to (3.1), we have:\psp

{\bf Theorem 3.2}. {\it Let $\be$ be any nonzero increasing
function of $t$ and let $\Im$ be any one-variable function.  We
have the following solutions of the two-dimensional classical
non-steady boundary layer problem (1.1) and (1.2): (1)
$$u=\frac{{\be'}'x}{2\be'}+
\frac{e^\be}{2\sqrt{\be'}}
\Im\left(\frac{x}{\sqrt{\be'}}\right)(e^{\sqrt{\be'}\:y}+a
e^{-\sqrt{\be'}\:y}), \eqno(3.42)$$
$$v=-\frac{{\be'}'y}{2\be'}-
\frac{e^\be}{2\sqrt{(\be')^3}}
\Im'\left(\frac{x}{\sqrt{\be'}}\right)(e^{\sqrt{\be'}\:y}-a
e^{-\sqrt{\be'}\:y})\eqno(3.43)$$ and $p$ is given in (3.40) with
$r=0$, where $a\in\mbb{R}$; (2)
$$u=\frac{{\be'}'x}{2\be'}+
\frac{e^{-\be}}{\sqrt{\be'}}
\Im\left(\frac{x}{\sqrt{\be'}}\right)\cos(\sqrt{\be'}\:y),
\eqno(3.44)$$
$$v=-\frac{{\be'}'y}{2\be'}-
\frac{e^{-\be}}{\sqrt{(\be')^3}}
\Im'\left(\frac{x}{\sqrt{\be'}}\right)\sin(\sqrt{\be'}\:y)
\eqno(3.45)$$ and $p$ is given in (3.40) with $r=1$.}\psp

Finally, we look for $\xi$ of the following form:
$$\xi=\kappa(t,x)y+g(t,x)+h(t,x)y^{-1},\eqno(3.46)$$
where $\kappa,g,h$ are functions of $t,x$ with $h\not\equiv 0$.
Then
$$\xi_y=\kappa-hy^{-2},\qquad \xi_{yt}=\kappa_t-h_ty^{-2},\qquad
\xi_{yx}=\kappa_x-h_xy^{-2},\eqno(3.47)$$
$$\xi_x=\kappa_xy+g_x+h_xy^{-1},\qquad\xi_{yy}=2hy^{-3},\qquad
\xi_{yyy}=-6hy^{-4}.\eqno(3.48)$$ So (3.2) becomes
$$\kappa_t-h_ty^{-2}+(\kappa-hy^{-2})(\kappa_x-h_xy^{-2})-2hy^{-3}(\kappa_xy+g_x+h_xy^{-1})
+p_x=-6hy^{-4},\eqno(3.49)$$ equivalently,
$$\kappa_t+\kappa\kappa_x+p_x-(h_t+\kappa h_x+3\kappa_xh)y^{-2}-2hg_xy^{-3}-hh_xy^{-4}=-6hy^{-4}.
\eqno(3.50)$$ Moreover, (3.50) is implied by the following
equations:
$$\kappa_t+\kappa\kappa_x+p_x=0,\qquad h_t+\kappa h_x+3\kappa_xh=0,
\eqno(3.51)$$
$$g=0,\qquad h_x=6.\eqno(3.52)$$
So $h=6x$ modulo $T_{4\al}$ in (2.5).

According to the second equation in (3.51),
$$6\kappa+18x\kappa_x=0\lra
\kappa=\gm x^{-1/3}\eqno(3.53)$$ for some function $\gm$ of $t$.
Thus
$$\xi=\gm x^{-1/3}y+6xy^{-1}. \eqno(3.54)$$ Moreover,
$\kappa_t=\gm' x^{-1/3}.$ Thus
$$p_x=\frac{\gm^2}{3}x^{-5/3}-\gm' x^{-1/3}\lra
p=\frac{3\gm'}{2} x^{2/3}-\frac{\gm^2}{2}x^{-2/3}\eqno(3.55)$$
 by modulo $T_{5\be}$ in
(2.6) by the first equation in (3.51). Therefore, (3.1)
implies:\psp

{\bf Theorem 3.3}. {\it Let $\gm$ be any function of $t$. We have
the following solutions of the two-dimensional classical
non-steady boundary layer problem (1.1) and (1.2):
$$u=\gm x^{-1/3}-6xy^{-2},\;\;
v=\frac{\gm}{3}x^{-4/3}-6y^{-1},\;\; p=\frac{3\gm'}{2}
x^{2/3}-\frac{\gm^2}{2}x^{-2/3}.\eqno(3.56)$$}

\section{3-D Uniform Exponential Approach}

In this section, we will find certain function-parameter exact
solutions with single exponential function in $y$ for the
three-dimensional non-steady boundary layer equations (1.3)-(1.5).

Solving the second equation in (1.5), we get
$$u=\xi_y,\qquad w=\eta_y,\qquad v=-(\xi_x+\eta_z)\eqno(4.1)$$
 for some functions $\xi$ and $\eta$ in $t,x,y,z$. Now the equation (1.3)
 and (1.4)
  become
$$\xi_{yt}+\xi_y\xi_{yx}-(\xi_x+\eta_z)\xi_{yy}+\eta_y\xi_{yz}
+\frac{1}{\rho}p_x= \nu\xi_{yyy},\eqno(4.2)$$
$$\eta_{yt}+\xi_y\eta_{yx}-(\xi_x+\eta_z)\eta_{yy}+\eta_y\eta_{yz}
+\frac{1}{\rho} p_z= \nu\eta_{yyy}.\eqno(4.3)$$

First we assume that  $\xi$ and $\eta $ are of the form:
$$\xi=f(t,x,z)+\kappa(t,x,z)y+g(t,x,z)H(t,\sgm(t,x,z)y),\eqno(4.4)$$
and
$$\eta=\tau(t,x,z)y+\zeta(t,x,z)
\Phi(t,\sgm(t,x,z)y),\eqno(4.5)$$ where
$f,g,\kappa,\sgm,\tau,\zeta$ are three-variable functions, and
$H(t,\varpi)$ and $\Phi(t,\varpi)$ are two variable functions,
where $\varpi=\sgm y$. Then
$$\xi_y=\kappa+\sgm gH_\varpi,\qquad \eta_y=\tau+\sgm \zeta\Phi_\varpi,
\eqno(4.6)$$
$$\xi_x=f_x+\kappa_xy+g_xH+\sgm_xgyH_\varpi,\qquad
\eta_z=\tau _zy+\zeta_z\Phi+\sgm_z\zeta\Phi_\varpi,\eqno(4.7)$$
$$\xi_{yx}=\kappa_x+(\sgm g)_xH_\varpi
+\sgm\sgm_xgyH_{\varpi\varpi},
\qquad\eta_{yx}=\tau_x+(\sgm\zeta)_x\Phi_\varpi
+\sgm\sgm_xgy\Phi_{\varpi\varpi},\eqno(4.8)$$
$$\xi_{yz}=\kappa_z+(\sgm g)_zH_\varpi
+\sgm\sgm_zgyH_{\varpi\varpi},
\qquad\eta_{yz}=\tau_z+(\sgm\zeta)_z\Phi_\varpi
+\sgm\sgm_zgy\Phi_{\varpi\varpi},\eqno(4.9)$$
$$\xi_{yy}=\sgm^2gH_{\varpi\varpi},\qquad\xi_{yyy}=\sgm^3gH_{\varpi\varpi\varpi},\qquad
\eta_{yy}=\sgm^2\zeta\Phi_{\varpi\varpi}, \qquad
\eta_{yyy}=\sgm^3\zeta\Phi_{\varpi\varpi\varpi}.\eqno(4.10)$$

Thus the nonlinear term in (4.2) becomes
\begin{eqnarray*}& &\xi_y\xi_{yx}-(\xi_x+\eta_z)\xi_{yy}+\eta_y\xi_{yz}
 =(\kappa+\sgm gH_\varpi)(\kappa_x+(\sgm g)_xH_\varpi
 +\sgm\sgm_xgyH_{\varpi\varpi})
\\ & &-\sgm^2gH_{\varpi\varpi}(f_x+(\kappa_x+\tau_z)y+g_xH
+\sgm_xgyH_\varpi+\zeta_z\Phi+\sgm_z\zeta y\Phi_\varpi )
\\ & &+(\tau+\sgm\zeta\Phi_\varpi) (\kappa_z+(\sgm
g)_zH_\varpi+\sgm\sgm_zgyH_{\varpi\varpi})\\
&=&\kappa\kappa_x+\tau\kappa_z+[(\sgm g\kappa)_x+(\sgm
g)_z\tau]H_\varpi +\sgm\zeta\kappa_z\Phi_\varpi+\sgm
g\{[\kappa\sgm_x+\tau\sgm_z -\sgm(\kappa_x+\tau_z)]y\\
& &-\sgm f_x\}H_{\varpi\varpi} +\sgm g(\sgm
 g)_xH_\varpi^2-\sgm^2 gg_xHH_{\varpi\varpi}+
 \sgm\zeta(\sgm g)_zH_\varpi\Phi_\varpi-\sgm^2
 g\zeta_zH_{\varpi\varpi}\Phi.\hspace{0.6cm}(4.11)
\end{eqnarray*}
Symmetrically, the nonlinear term in (4.3) becomes
\begin{eqnarray*}& &\xi_y\eta_{yx}-(\xi_x+\eta_z)\eta_{yy}
+\eta_y\eta_{yz}\\
&=&\kappa\tau_x+\tau\tau_z+[(\sgm \zeta\tau)_z+(\sgm
\zeta)_x\kappa]\Phi_\varpi +\sgm g\tau_xH_\varpi+\sgm
\zeta\{[\kappa\sgm_x+\tau\sgm_z -\sgm(\kappa_x+\tau_z)]y\\
& &-\sgm f_x\}\Phi_{\varpi\varpi} +\sgm \zeta(\sgm
 \zeta)_z\Phi_\varpi^2-\sgm^2 \zeta \zeta_z\Phi\Phi_{\varpi\varpi}+
 \sgm g(\sgm \zeta)_xH_\varpi\Phi_\varpi-\sgm^2
 g_x\zeta H\Phi_{\varpi\varpi}.\hspace{0.8cm}(4.12)
\end{eqnarray*}

In order to linearize (4.11) and (4.12) with respect to $H$ and
$\Phi$, we divide it into ten cases. \vspace{0.4cm}

{\it Case 1}. $H=\Phi=e^\varpi$.\psp

 In this case,
\begin{eqnarray*}& &\xi_y\xi_{yx}-(\xi_x+\eta_z)\xi_{yy}+\eta_y\xi_{yz}
=\kappa\kappa_x+\tau\kappa_z+\{\sgm g\kappa_x+\sgm\zeta
\kappa_z+(\sgm g)_x\kappa+(\sgm g)_z\tau\\ & &+\sgm
g\{[\kappa\sgm_x+\tau\sgm_z -\sgm(\kappa_x+\tau_z)]y-\sgm
f_x\}\}e^\varpi
 +\sgm[\sgm_xg^2+\zeta(\sgm g)_z-\sgm
 g\zeta_z]e^{2\varpi}.\hspace{1cm}(4.13)
\end{eqnarray*}
Moreover,
$$\xi_{yt}=\kappa_t+((\sgm
g)_t+\sgm\sgm_tgy)e^\varpi,\qquad\xi_{yyy}=\sgm^3
ge^\varpi.\eqno(4.14)$$ Thus (4.2) becomes
\begin{eqnarray*}& &\kappa_t+\kappa\kappa_x+\tau\kappa_z+\{(\sgm
g)_t+\sgm g\kappa_x+\sgm\zeta \kappa_z+(\sgm g)_x\kappa+(\sgm
g)_z\tau+\sgm g\{[\sgm_t+\kappa\sgm_x+\tau\sgm_z
\\
& &-\sgm(\kappa_x+\tau_z)]y-\sgm f_x-\nu\sgm^2\}\}e^\varpi +\sgm
[\sgm_xg^2+\zeta(\sgm g)_z-\sgm
 g\zeta_z]e^{2\varpi}+\frac{1}{\rho}p_x=0.\hspace{1cm}(4.15)
\end{eqnarray*}
By symmetry, (4.2) and (4.3) are implied by the following system
of partial differential equations:
$$\kappa_t+\kappa\kappa_x+\tau\kappa_z+\frac{1}{\rho}p_x=0,\qquad
\tau_t+\kappa\tau_x+\tau\tau_z+\frac{1}{\rho}p_z=0,\eqno(4.16)$$
$$\sgm_t+\kappa\sgm_x+\tau\sgm_z-\sgm(\kappa_x+\tau_z)=0,\eqno(4.17)$$
$$(\sgm g)_t+\sgm g\kappa_x+\sgm\zeta \kappa_z+(\sgm
g)_x\kappa+(\sgm g)_z\tau-\sgm^2 g( f_x+\nu\sgm)=0,\eqno(4.18)$$
$$(\sgm\zeta)_t+\sgm g\tau_x+\sgm\zeta \tau_z+(\sgm
\zeta)_x\kappa+(\sgm\zeta)_z\tau-\sgm^2 \zeta(
f_x+\nu\sgm)=0,\eqno(4.19)$$
$$\sgm_xg^2+\zeta(\sgm g)_z-\sgm
 g\zeta_z=0,\qquad\sgm_z\zeta^2+g(\sgm \zeta)_x-\sgm
 g_x\zeta=0
.\eqno(4.20)$$

 Observe that (4.20) yields
$$
\sgm_x+\sgm_z\frac{\zeta}{g}=\sgm\left(\frac{\zeta}{g}\right)_z,\qquad
\sgm_z+\sgm_x\frac{g}{\zeta}=\sgm\left(\frac{g}{\zeta}\right)_x
.\eqno(4.21)$$ Thus we take
$$\left(\frac{\zeta}{g}\right)_z=\left(\frac{\zeta}{g}\right)\left(\frac{g}{\zeta}\right)_x
\lra\left(\frac{\zeta}{g}\right)\left(\frac{\zeta}{g}\right)_z=
-\left(\frac{\zeta}{g}\right)_x.\eqno(4.22)$$
 Modulo
$T_{5\al},T_{6\al}$ in (2.17)-(2.20), we take the solutions:
$$\frac{\zeta}{g}=\tan\gm,\;\frac{z}{x},\eqno(4.23)$$
where $\gm$ is a function of $t$. In the first case,
$$\sgm_x+\sgm_z\tan\gm=0,\eqno(4.24)$$
and we take $$\sgm=e^{\Im(z\cos\gm-x\sin\gm)}\sec\gm\eqno(4.25)$$
for later convenience. In the second case,
$$x\sgm_x+z\sgm_z=\sgm,\eqno(4.26)$$
and we take $$\sgm=xe^{\Im(x/z)+\gm_1},\eqno(4.27)$$ where $\gm_1$
is a function of $t$ and $\Im$ is a one-variable function.
Furthermore, $\sgm g\times (4.19)-\sgm\zeta\times(4.18)$ gives
\begin{eqnarray*}& &(\sgm\zeta)_t\sgm g-\sgm\zeta(\sgm g)_t+
(\sgm g)^2\tau_x -(\sgm\zeta)^2\kappa_z+(\sgm\zeta)(\sgm
g)(\tau_z-\kappa_x)\\& &+[(\sgm\zeta)_x\sgm g-\sgm\zeta(\sgm
g)_x]\kappa+[(\sgm\zeta)_z\sgm g-\sgm\zeta(\sgm
g)_z]\tau=0.\hspace{4.9cm}(4.28)\end{eqnarray*} Multiplying (4.28)
by $1/(\sgm g)^2$, we obtain
$$\left(\frac{\zeta}{g}\right)_t+\tau_x
-\left(\frac{\zeta}{g}\right)^2\kappa_z+
\left(\frac{\zeta}{g}\right)(\tau_z-\kappa_x)+
\left(\frac{\zeta}{g}\right)_x\kappa
+\left(\frac{\zeta}{g}\right)_z\tau=0.\eqno(4.29)$$ Since
$p_{xz}=p_{xz}$ by our convention, (4.16) implies
$$(\kappa_z-\tau_x)_t+(\kappa(\kappa_z-\tau_x))_x
+(\tau(\kappa_z-\tau_x))_z=0.\eqno(4.30)$$

Assume $\zeta/g=\tan \gm$. Denote the  moving frame
$$\td\varpi=x\cos\gm+z\sin\gm,\qquad \hat\varpi=z\cos\gm-x\sin\gm.\eqno(4.31)$$
Then
$$\ptl_{\td\varpi}=\cos\gm\:\ptl_x+\sin\gm\:\ptl_z,\qquad
\ptl_{\hat\varpi}=-\sin\gm\:\ptl_x+\cos\gm\:\ptl_z.\eqno(4.32)$$
Observe that (4.29) becomes
$$\frac{\gm'}{\cos^2\gm}+\tau_x-\kappa_z\tan^2\gm
+(\tau_z-\kappa_x)\tan\gm=0,\eqno(4.33)$$ equivalently,
$$\gm'+(\cos\gm\:\ptl_x+\sin\gm\;\ptl_z)(\tau\cos\gm-\kappa\sin\gm)=0
\sim\gm'+
\ptl_{\td\varpi}(\tau\cos\gm-\kappa\sin\gm)=0,\eqno(4.34)$$ and so
we take
$$\tau\cos\gm-\kappa\sin\gm=-\gm'\td\varpi\lra \tau=\kappa\tan\gm-
\gm'\td\varpi\sec\gm \eqno(4.35)$$ for the convenience of solving
the problem. Note
$$\ptl_t(\td\varpi)=\gm'\hat\varpi,\qquad
\ptl_t(\hat\varpi)=-\gm'\td\varpi.\eqno(4.36)$$ Substituting
(4.25) and (4.35) into (4.17), we obtain
$$2\gm'\sin\gm-2\gm'\cos\gm\:\td\varpi\Im'
-\ptl_{\td\varpi}(\kappa)=0,\eqno(4.37)$$ where $\Im$ is a
function of $\hat\varpi$. Thus we take
$$\kappa=2\gm'\td\varpi\sin\gm-\gm'
\td\varpi^2\Im'\cos\gm+\ves(t,\hat\varpi)
 \eqno(4.38)$$ for some two-variable function
$\ves(t,\hat\varpi)$. Moreover,
$$\tau=-\gm'\td\varpi\frac{\cos 2\gm}{\cos\gm}-\gm'\td\varpi^2\Im'
\sin\gm+\ves\tan\gm.\eqno(4.39)$$

Note
$$\kappa_z-\tau_x=-\gm'
\td\varpi^2{\Im'}'+\ves_{\hat\varpi}\sec\gm+\gm'. \eqno(4.40)$$
Moreover,
\begin{eqnarray*}(\kappa_z-\tau_x)_t&=&-{\gm'}'
\td\varpi^2{\Im'}'-2(\gm')^2 \td\varpi\hat\varpi{\Im'}'+(\gm')^2
\td\varpi^3{{\Im'}'}'\\ & & +\sec\gm\;(\ves_{\hat\varpi
t}+\gm'\tan\gm\:\ves_{\hat\varpi}-\gm'\td\varpi
\ves_{\hat\varpi\hat\varpi})+ {\gm'}',
\hspace{4.5cm}(4.41)\end{eqnarray*}
\begin{eqnarray*}&
&[\kappa(\kappa_z-\tau_x)]_x+[\tau(\kappa_z-\tau_x)]_z=(\gm')^2
\left[4\Im'{\Im'}'+{{\Im'}'}' \right] \td\varpi^3
-3(\gm')^2\tan\gm\\ & &\times{\Im'}'\td\varpi^2 -\gm'
[\sec\gm(2\ves\Im'+\ves_{\hat\varpi})+
 2\gm'\Im]_{\hat\varpi}\td\varpi
 +\gm'\tan\gm\;(\sec\gm\:\ves_{\hat\varpi}+\gm').\hspace{2.3cm}(4.42)
 \end{eqnarray*}
Since the subcase $\gm\in\mbb{R}$ is included in Case 3 modulo the
transformation in (2.10)-(2.12), we assume $\gm'\not\equiv 0$.
 Thus   (4.30) is implied by
the following system of partial differential equations:
$${{\Im'}'}'+2\Im'{\Im'}'=0,\qquad
({\gm'}'+3(\gm')^2\tan\gm){\Im'}'=0,\eqno(4.43)$$
$${\gm'}'+(\gm')^2\tan\gm+2\gm'\ves_{\hat\varpi}\sec\gm\;\tan\gm+
\ves_{\hat\varpi t}\sec\gm=0, \eqno(4.44)$$
$$[(\ves\Im'+\ves_{\hat\varpi})\sec\gm+
 \gm'\hat\varpi\Im']_{\hat\varpi}
=0. \eqno(4.45)$$

 By (4.44), we take
$$\ves=\iota(\hat\varpi)\cos^2\gm-\gm'\hat\varpi\cos\gm\eqno(4.46)$$
for some function $\iota$ of $\hat\varpi$.
 Moreover, (4.45) is implied by
$$(\Im'(\hat\varpi)\iota(\hat\varpi)
+\iota'(\hat\varpi))_{\hat\varpi} =0.\eqno(4.47)$$ First,
$$\gm\;\;\mbox{is arbitrary and}\;\;\iota=c_1+d_1e^{-a\hat\varpi}\qquad
\mbox{if}\;\;\Im=a\hat\varpi+b, \eqno(4.48)$$ where
$a,b,c_1,d_1\in\mbb{R}$. Assume ${\Im'}'\neq 0$, we have
$${{\Im'}'}'(\hat\varpi)+2\Im'(\hat\varpi){\Im'}'(\hat\varpi)=0,\qquad
{\gm'}'+3(\gm')^2\tan\gm=0.\eqno(4.49)$$ Thus
$$\Im=\ln (a\hat\varpi+b),\;
\ln \sin(a\hat\varpi+b),\;\ln
(be^{a\hat\varpi}+b_1e^{-a\hat\varpi})\eqno(4.50)$$ with
$a,b,b_1\in\mbb{R}$ and $\gm$ is implicitly given by
$$\frac{1+\sin\gm}{1-\sin\gm}=ce^{dt-2\sec\gm\:\tan\gm},\qquad
c,d\in\mbb{R}.\eqno(4.51)$$ Furthermore,
$$\iota=c_1(a\hat\varpi+b)+\frac{d_1}{a\hat\varpi+b}\qquad\mbox{if}\;\;
\Im=\ln(a\hat\varpi+b),\eqno(4.52)$$
$$\iota=c_1\cot(a\hat\varpi+b)+d_1\csc
(a\hat\varpi+b)\qquad\mbox{if}\;\;\Im=\ln\sin(a\hat\varpi+b),\eqno(4.53)$$
$$\iota=c_1\frac{be^{a\hat\varpi}-b_1e^{-a\hat\varpi}}
{be^{a\hat\varpi}+b_1e^{-a\hat\varpi}}+\frac{d_1}
{be^{a\hat\varpi}+b_1e^{-a\hat\varpi}}\qquad\mbox{if}\;\;\Im=\ln
(be^{a\hat\varpi}+b_1e^{-a\hat\varpi}).\eqno(4.54)$$

According to (4.35), (4.38) and (4.39),
\begin{eqnarray*}&&\kappa_t+\kappa\kappa_x+\tau\kappa_z\\ &=&
\kappa_t+\kappa\ptl_{\td\varpi}(\kappa)\sec\gm-\gm'\td\varpi
\kappa_z\sec\gm\\
&=&2({\gm'}'\sin\gm+(\gm')^2\cos\gm)\td\varpi+
2(\gm')^2\hat\varpi\sin\gm+\ves_t-\gm'\td\varpi \ves_{\hat\varpi}
\hspace{5cm}\end{eqnarray*}
\begin{eqnarray*}
&&+((\gm')^2\sin\gm-{\gm'}'\cos\gm) \td\varpi^2\Im'-2(\gm')^2
\hat\varpi\td\varpi\Im'\cos\gm+(\gm')^2 \td\varpi^3{\Im'}'\cos\gm
\\& &+(2\gm'\td\varpi\sin\gm-\gm' \td\varpi^2\Im'\cos\gm+\ves)
(2\gm'\tan\gm-2\gm' \td\varpi\Im')
\\ & & -\gm'
\td\varpi(2\gm'\sin^2\gm\;\sec\gm-\gm'
(2\td\varpi\Im'\sin\gm+\td\varpi^2{\Im'}'\cos\gm)+\ves_{\hat\varpi})
\\ &=&2(\gm')^2 \td\varpi^3({\Im'}'+(\Im')^2)\cos\gm-
(3(\gm')^2\sin\gm+{\gm'}'\cos\gm)
\td\varpi^2\Im'+\ves_t+2\gm'(\ves\tan\gm\\ &
&+\gm'\hat\varpi\sin\gm)
+2({\gm'}'\sin\gm+(\gm')^2\sec\gm-\gm'(\gm'\hat\varpi\Im'\cos\gm+
\ves_{\hat\varpi}+\Im'\ves)) \td\varpi,\hspace{1.4cm}(4.55)
\end{eqnarray*}
\begin{eqnarray*}&&\tau_t+\kappa\tau_x+\tau\tau_z\\ &=&
\tau_t+\kappa\ptl_{\td\varpi}(\tau)\sec\gm-\gm'\td\varpi
\tau_z\sec\gm\\
&=&\left(4(\gm')^2\sin\gm-(\gm')^2\frac{\cos
2\gm\:\sin\gm}{\cos^2\gm}-{\gm'}'\frac{\cos 2\gm}{\cos\gm}\right)
\td\varpi-(\gm')^2\frac{\cos 2\gm}{\cos\gm}\hat\varpi -2(\gm')^2
\hat\varpi\td\varpi\Im'\sin\gm\\ & &-({\gm'}'
\sin\gm+(\gm')^2\cos\gm)\td\varpi^2\Im'+(\gm')^2\td\varpi^3{\Im'}'\sin\gm
+\gm'\sec^2\gm\;\ves+
(\ves_t-\gm'\td\varpi\;\ves_{\hat\varpi})\tan\gm
\\ &&-\gm'(2\gm'\td\varpi\sin\gm-\gm' \td\varpi^2\Im'\cos\gm+\ves)
\left(\frac{\cos 2\gm}{\cos^2\gm}+2\td\varpi\Im' \tan\gm\right)
+\gm' \td\varpi[\gm'\cos2\gm\:\tan\gm
\\ & &+\gm'\td\varpi(2\Im'\sin^2\gm+\td\varpi{\Im'}'\sin\gm\:\cos\gm)
-\ves_{\hat\varpi}\sin\gm]\sec\gm\\ &=& 2(\gm')^2
\td\varpi^3({\Im'}'+(\Im')^2)\sin\gm-
(3(\gm')^2\sin\gm+{\gm'}'\cos\gm) \td\varpi^2\Im'\tan\gm\\&
&+2\gm'(\gm'\sec\gm-(\gm'\hat\varpi\Im'\cos\gm+
\ves_{\hat\varpi}+\Im'\ves))\td\varpi\tan\gm
\\ & &-\frac{\cos 2\gm}{\cos\gm}({\gm'}'\td\varpi+(\gm')^2\hat\varpi)
+(\ves_t+2\ves\tan\gm)\tan\gm.
 \hspace{5.5cm}(4.56)
\end{eqnarray*}
Note
$${\Im'}'+(\Im')^2=\hat b=\left\{\begin{array}{ll}a^2&\mbox{in Cases (4.48)
and (4.54),}\\
0&\mbox{in Case (4.52),}\\ -a^2&\mbox{in Case
(4.53),}\end{array}\right.\eqno(4.57)$$
$$\gm'\hat\varpi\Im'\cos\gm+
\ves_{\hat\varpi}+\Im'\ves=\iota'+\Im'\iota=\hat c=
\left\{\begin{array}{ll}ac_1&\mbox{in Cases (4.48) and (4.54)},\\
2ac_1&\mbox{in Cases (4.52)},\\ -ac_1&\mbox{in Case
(4.53),}\end{array}\right.\eqno(4.58)$$ and
$$\ves_t+2\ves\tan\gm=(\gm')^2\td\varpi\cos\gm-((\gm')^2\sin\gm+{\gm'}'\cos\gm)
\hat\varpi\eqno(4.59)$$ by (4.46). Moreover,
$$2\td\varpi\sin\gm-\hat\varpi\cos\gm=3x\sin\gm\:\cos\gm
+z(3\sin^2\gm-1),\eqno(4.60)$$
$$-\frac{\cos 2\gm}{\cos\gm}\td\varpi-\hat\varpi\sin\gm=
x(3\sin^2\gm-1)+z(1-3\cos^2\gm)\tan\gm,\eqno(4.61)$$
$$-\frac{\cos
2\gm}{\cos\gm}-\sin\gm\:\tan\gm=-\cos\gm.\eqno(4.62)$$ By (4.16),
(4.43) and (4.56)-(4.63), we have:
\begin{eqnarray*}p&=&\rho\{\frac{3(\gm')^2\tan\gm+{\gm'}'}{3}\td\varpi^3\Im'
+\frac{(\gm')^2 [\hat\varpi^2-(\hat
b\td\varpi^2+1)\td\varpi^2]}{2}+ \gm'\td\varpi^2(\hat
c-\gm\sec\gm)\sec\gm
\\& & +{\gm'}'\left(xz(1-3\sin^2\gm)-\frac{3x^2\sin2\gm}{4}
+\frac{z^2(3\cos^2\gm-1)\tan\gm}{2}\right)\}\hspace{2.7cm}(4.63)
\end{eqnarray*}
modulo the transformation in (1.26).

Modulo the transformation given in (2.22) and (2.23), we take
$$g=\sgm^{-1}=\cos\gm\:e^{-\Im(\hat\varpi)}\eqno(4.64)$$
by (4.25). According to (4.18), (4.23) and (4.38), we have
$$f_x=2\gm'e^{-\Im}(
\sin\gm-\td\varpi\Im'\cos\gm) -\nu e^{\Im}\sec\gm.\eqno(4.65)$$
Furthermore,
$$\xi=f+(2\gm'\td\varpi\sin\gm-\gm'\td\varpi^2\Im'
\cos\gm+\ves)y+\cos\gm\: \exp(e^{\Im}y\sec\gm-\Im),\eqno(4.66)$$
$$\eta=\left(\ves\tan\gm-\gm'\frac{\cos 2\gm}{\cos\gm}\td\varpi-\gm'
\td\varpi^2\Im'\sin\gm\right)y+\sin\gm\: \exp(\sec\gm\;e^\Im
y-\Im).\eqno(4.67)$$
 By (4.1), we have the following
theorem. \psp

{\bf Theorem 4.1}. {\it In terms of the notations in (4.31),  we
have the following solutions of the three-dimensional classical
non-steady boundary layer equations (1.3), (1.4) and (1.5):
$$u=2\gm'\td\varpi\sin\gm-\gm'
\td\varpi^2\Im'\cos\gm+\ves+ \exp(e^{\Im}y\sec\gm),\eqno(4.68)$$
$$w=\ves\tan\gm-\gm'\frac{\cos 2\gm}{\cos\gm}\td\varpi-\gm'
\td\varpi^2\Im'\sin\gm+\tan\gm\;\exp(e^{\Im}y\sec\gm),\eqno(4.69)$$
$$v=2\gm'e^{-\Im}(
\cos\gm\;\td\varpi\Im'-\sin\gm)+\nu e^{\Im}\sec\gm
+\gm'(2\td\varpi\Im'-\tan\gm)y\eqno(4.70)$$
 and $p$ is given in (4.63), where (1)(4.48) is taken and
 $\ves$ is given (4.46); (2)
$\gm$ is implicitly determined by (4.51), (4.52)-(4.54) are taken
and $\ves$ is given (4.46).}\psp

Next we consider the case $\zeta/g=z/x$. Recall that we have
already determined $\sgm$ in (4.27). Now (4.29) becomes
$$x^2\tau_x
-z^2\kappa_z+xz(\tau_z-\kappa_x) -z\kappa +x\tau=0.\eqno(4.71)$$
equivalently,
$$(x\ptl_x+z\ptl_z+1)\left(\tau-\frac{z}{x}\kappa\right)=0.\eqno(4.72)$$
Thus
$$\tau=\frac{z}{x}\kappa+x^{-1}\ves(t,x/z)\eqno(4.73)$$
for some two-variable function $\ves$. Denote
$$\hat\varpi=\frac{x}{z}.\eqno(4.74)$$
 Substituting (4.27) into
(4.17), we get
$$x\gm_1'+\kappa(1+\hat\varpi\Im')
-\tau\hat\varpi^2\Im'-x(\kappa_x+\tau_z)=0.\eqno(4.75)$$ By
(4.73), it changes to
$$x\kappa_x+z\kappa_z=x\gm_1'+x^{-1}\hat\varpi^2(\ves_{\hat\varpi}
-\ves\Im'). \eqno(4.76)$$ Thus
$$\kappa=x\gm_1'+x^{-1}\hat\varpi^2(\ves\Im'
-\ves_{\hat\varpi})+\frac{\hat\varpi}{\sqrt{1+\hat\varpi^2}}
\vt(t,\hat\varpi),\eqno(4.77)$$ where $\vt$ is another
two-variable function and the last terms is so written for later
convenience. Moreover,
$$\tau=z\gm_1'+
x^{-1}\hat\varpi(\ves\Im'-\ves_{\hat\varpi})
+\frac{\vt}{\sqrt{1+\hat\varpi^2}}+x^{-1}\ves.\eqno(4.78)$$

For convenience of solving the problem, we denote
$$\phi(t,\hat\varpi)=\hat\varpi^2(\ves\Im'-\ves_{\hat\varpi}),\qquad
\psi(t,\hat\varpi)=\hat\varpi^{-2}\phi+\hat\varpi^{-1}\ves.
\eqno(4.79)$$ So
$$\kappa=x\gm_1'+x^{-1}\phi+\frac{\hat\varpi\vt}{\sqrt{1+\hat\varpi^2}},
\qquad\tau=z\gm_1'+z^{-1}\psi+
\frac{\vt}{\sqrt{1+\hat\varpi^2}}.\eqno(4.80)$$ Observe that
$$\kappa_z=-x^{-2}\hat\varpi^2\phi_{\hat\varpi}
-x^{-1}\left(\frac{\hat\varpi^2(\vt+\hat\varpi\vt_{\hat\varpi})}
{\sqrt{1+\hat\varpi^2}}
-\frac{\hat\varpi^4\vt}{\sqrt{(1+\hat\varpi^2)^3}}
\right),\eqno(4.81)$$
$$\tau_x=x^{-2}\hat\varpi^2\psi_{\hat\varpi}+
x^{-1}\left(\frac{\hat\varpi\vt_{\hat\varpi}}{\sqrt{1+\hat\varpi^2}}
-\frac{\hat\varpi^2\vt}{\sqrt{(1+\hat\varpi^2)^3}}\right).\eqno(4.82)$$
Thus
$$\tau_x-\kappa_z=x^{-2}\hat\varpi^2(\psi+\phi)_{\hat\varpi}
+x^{-1}\hat\varpi\sqrt{1+\hat\varpi^2}\:\vt_{\hat\varpi}\eqno(4.83)$$
and
$$(\tau_x-\kappa_z)_t=x^{-2}\hat\varpi^2(\psi+\phi)_{t\hat\varpi}
+x^{-1}\hat\varpi\sqrt{1+\hat\varpi^2}\:\vt_{t\hat\varpi}
.\eqno(4.84)$$ Moreover,
\begin{eqnarray*}\kappa(\tau_x-\kappa_z)&=&
x^{-3}\hat\varpi^2\phi(\psi+\psi)_{\hat\varpi}+x^{-2}
\left[\frac{\hat\varpi^3\vt}{\sqrt{1+\hat\varpi^2}}
(\psi+\phi)_{\hat\varpi}+
\hat\varpi\sqrt{1+\hat\varpi^2}\:\phi\vt_{\hat\varpi}\right]\\ &
&+
x^{-1}\hat\varpi^2[\gm_1'(\psi+\phi)_{\hat\varpi}+\vt\vt_{\hat\varpi}]
+\gm_1'\hat\varpi\sqrt{1+\hat\varpi^2}\:\vt_{\hat\varpi}
\hspace{3.7cm}(4.85)\end{eqnarray*} and
\begin{eqnarray*}\tau(\tau_x-\kappa_z)&=&
x^{-3}\hat\varpi^3\psi(\psi+\psi)_{\hat\varpi}+x^{-2}
\left[\frac{\hat\varpi^2\vt}{\sqrt{1+\hat\varpi^2}}
(\psi+\phi)_{\hat\varpi}+
\hat\varpi^2\sqrt{1+\hat\varpi^2}\:\psi\vt_{\hat\varpi}\right]\\ &
&+
x^{-1}\hat\varpi[\gm_1'(\psi+\phi)_{\hat\varpi}+\vt\vt_{\hat\varpi}]
+\gm_1'\sqrt{1+\hat\varpi^2}\:\vt_{\hat\varpi}.
\hspace{4.1cm}(4.86)\end{eqnarray*}

Observe that
\begin{eqnarray*}& &[\kappa(\tau_x-\kappa_z)]_x=
x^{-4}\hat\varpi^2(\hat\varpi\ptl_{\hat\varpi}-1)
(\phi(\psi+\psi)_{\hat\varpi})+x^{-1}\gm_1'\hat\varpi(1+\hat\varpi\ptl_{\hat\varpi})
(\sqrt{1+\hat\varpi^2}\vt_{\hat\varpi})\\ & &+x^{-3}
\left[\hat\varpi^3(1+\hat\varpi\ptl_{\hat\varpi})\left(\frac{\vt}{\sqrt{1+\hat\varpi^2}}
(\psi+\phi)_{\hat\varpi}\right)+
\hat\varpi(\hat\varpi\ptl_{\hat\varpi}-1)(\sqrt{1+\hat\varpi^2}
\phi\vt_{\hat\varpi})\right]\\
& &+ x^{-2}\hat\varpi^2(1+\hat\varpi\ptl_{\hat\varpi})
[\gm_1'(\psi+\phi)_{\hat\varpi}+\vt\vt_{\hat\varpi}]
\hspace{7.4cm}(4.87)\end{eqnarray*} and
\begin{eqnarray*}& &[\tau(\tau_x-\kappa_z)]_z=
-x^{-4}\hat\varpi^4(\hat\varpi\ptl_{\hat\varpi}+3)
(\psi(\psi+\psi)_{\hat\varpi})-x^{-1}
\gm_1'\hat\varpi^2\ptl_{\hat\varpi}(\sqrt{1+\hat\varpi^2}\vt_{\hat\varpi})
\\ & &
-x^{-3} \hat\varpi^3(\hat\varpi\ptl_{\hat\varpi}+2)
\left[\frac{\vt}{\sqrt{1+\hat\varpi^2}} (\psi+\phi)_{\hat\varpi}+
\sqrt{1+\hat\varpi^2}\psi\vt_{\hat\varpi}\right]\\
&
&-x^{-2}\hat\varpi^2(\hat\varpi\ptl_{\hat\varpi}+1)[\gm_1'(\psi+\phi)_{\hat\varpi}+\vt\vt_{\hat\varpi}]
. \hspace{7.3cm}(4.88)\end{eqnarray*} By (4.84), (4.87) and
(4.88), (4.30) is equivalent to the following system of partial
differential equations:
$$(\hat\varpi\ptl_{\hat\varpi}-1)
(\phi(\psi+\psi)_{\hat\varpi})=\hat\varpi^2(\hat\varpi\ptl_{\hat\varpi}+3)
(\psi(\psi+\psi)_{\hat\varpi}),\eqno(4.89)$$
$$(\hat\varpi\ptl_{\hat\varpi}-1)(\sqrt{1+\hat\varpi^2}
\phi\vt_{\hat\varpi})-\hat\varpi^2(\hat\varpi\ptl_{\hat\varpi}+2)
(\sqrt{1+\hat\varpi^2}\psi\vt_{\hat\varpi})-\frac{\hat\varpi^2\vt}{\sqrt{1+\hat\varpi^2}}
(\psi+\phi)_{\hat\varpi}=0,\eqno(4.90)$$
$$\vt_{t\hat\varpi}+\gm_1'\vt_{\hat\varpi}=0,\qquad
(\psi+\phi)_{t\hat\varpi}=0.\eqno(4.91)$$ According to (4.91),
$$\vt=\al(t)+e^{-\gm_1}\iota_1(\hat\varpi),\qquad
\psi+\phi=\be_1(t)+\iota(\hat\varpi),\eqno(4.92)$$ where
$\al,\be_1,\iota_1$ and $\iota$ are arbitrary one-variable
functions. Write (4.89)  as
$$\hat\varpi\ptl_{\hat\varpi}[\iota'(\phi-\hat\varpi^2\psi)]
=\iota'(\phi+\hat\varpi^2\psi).\eqno(4.93)$$ Moreover, (4.90) is
equivalent to
$$\hat\varpi(1+\hat\varpi^2)\ptl_{\hat\varpi}[\iota_1'
(\phi-\hat\varpi^2\psi)] -\iota_1'(\phi+\hat\varpi^4\psi)-
\hat\varpi^2(\al e^{\gm_1}+\iota_1)\iota'=0.\eqno(4.94)$$

 On the other
hand, the first equation in (4.79) yields
$$\ves=e^{\Im(\hat\varpi)}\hat\ves(t,\hat\varpi),\qquad
\phi=\hat\varpi^2e^{\Im(\hat\varpi)}\hat\ves_{\hat\varpi}(t,\hat\varpi),
\eqno(4.95)$$ where $\hat\ves$ is a function of $t$ and
$\hat\varpi$. According  the second equation in (4.79) and (4.92),
$$\phi+\hat\varpi^{-2}\phi+\hat\varpi^{-1}\ves
=\be_1+\iota \lra
e^\Im[(1+\hat\varpi^2)\hat\ves_{\hat\varpi}+\hat\varpi^{-1}\hat\ves]
=\be_1+\iota.\eqno(4.96)$$ Thus
$$\hat\ves=\frac{\sqrt{1+\hat\varpi^2}}{\hat\varpi}\left[\be+\be_1\int\frac{\hat\varpi
e^{-\Im}}{\sqrt{(1+\hat\varpi^2)^3}}\:d\hat\varpi+\int\frac{\hat\varpi\iota
e^{-\Im}}{\sqrt{(1+\hat\varpi^2)^3}}\:d\hat\varpi\right]\eqno(4.97)$$
for some function $\be$ of $t$. Thus
$$\phi=\frac{(\be_1+\iota)\hat\varpi^2}{1+\hat\varpi^2}-
\frac{e^\Im}{\sqrt{1+\hat\varpi^2}}\left[\be+
\be_1\int\frac{\hat\varpi
e^{-\Im}}{\sqrt{(1+\hat\varpi^2)^3}}\:d\hat\varpi+\int\frac{\hat\varpi\iota
e^{-\Im}}{\sqrt{(1+\hat\varpi^2)^3}}\:d\hat\varpi\right].\eqno(4.98)$$

Note
\begin{eqnarray*}& &\hat\varpi\ptl_{\hat\varpi}[\iota'(\phi-\hat\varpi^2
\psi)]
=\hat\varpi\ptl_{\hat\varpi}[\iota'((1+\hat\varpi^2)\phi
-(\be_1+\iota)\hat\varpi^2)]\\&=&-\hat\varpi\ptl_{\hat\varpi}\left\{
e^\Im\sqrt{1+\hat\varpi^2}\iota'\left[\be+
\be_1\int\frac{\hat\varpi
e^{-\Im}}{\sqrt{(1+\hat\varpi^2)^3}}\:d\hat\varpi+\int\frac{\hat\varpi\iota
e^{-\Im}}{\sqrt{(1+\hat\varpi^2)^3}}\:d\hat\varpi\right]\right\}
\\
&=&-\hat\varpi\ptl_{\hat\varpi}[e^\Im\sqrt{1+\hat\varpi^2}\iota']
\left[\be+ \be_1\int\frac{\hat\varpi
e^{-\Im}}{\sqrt{(1+\hat\varpi^2)^3}}\:d\hat\varpi+\int\frac{\hat\varpi\iota
e^{-\Im}}{\sqrt{(1+\hat\varpi^2)^3}}\:d\hat\varpi\right]
\\ & &-\frac{(\be_1+\iota)\iota'\hat\varpi^2}{1+\hat\varpi^2}
,\hspace{11cm}(4.99)
\end{eqnarray*}
\begin{eqnarray*}& &\iota'(\phi+\hat\varpi^2\psi)
=\iota'((1-\hat\varpi^2)\phi
+(\be_1+\iota)\hat\varpi^2)=\frac{2(\be_1+\iota)\iota'\hat\varpi^2}{1+\hat\varpi^2}
\\ & &-\frac{\iota'(1-\hat\varpi^2)e^\Im}{\sqrt{1+\hat\varpi^2}}\left[\be+
\be_1\int\frac{\hat\varpi
e^{-\Im}}{\sqrt{(1+\hat\varpi^2)^3}}\:d\hat\varpi+\int\frac{\hat\varpi\iota
e^{-\Im}}{\sqrt{(1+\hat\varpi^2)^3}}\:d\hat\varpi\right].\hspace{2cm}(4.100)\end{eqnarray*}
Thus (4.93) forces us to take $\iota'=0$.
 Replacing $\be_1$ by $\be_1+\iota$, we have $\iota=0$, and (4.93) naturally
 holds. Similarly, (4.94) yields $\iota_1=0$, and (4.94) naturally
 holds.

 According to (4.92), (4.95) and (4.98),
 $$\vt=\al,\qquad \phi=\be_1-\psi=
\frac{\be_1\hat\varpi^2}{1+\hat\varpi^2}-
\frac{e^\Im}{\sqrt{1+\hat\varpi^2}}\left(\be+
\be_1\int\frac{\hat\varpi
e^{-\Im}}{\sqrt{(1+\hat\varpi^2)^3}}\:d\hat\varpi\right).\eqno(4.101)$$
By (4.80),
$$\kappa=x\gm_1'+\be_1x^{-1}\left(\frac{\hat\varpi^2}{1+\hat\varpi^2}-
\frac{e^\Im}{\sqrt{1+\hat\varpi^2}}\int\frac{\hat\varpi
e^{-\Im}}{\sqrt{(1+\hat\varpi^2)^3}}\:d\hat\varpi\right)
+\frac{\al\hat\varpi-\be
x^{-1}e^\Im}{\sqrt{1+\hat\varpi^2}},\eqno(4.102)$$
$$\tau=z\gm_1'+\be_1z^{-1}\left(\frac{1}{1+\hat\varpi^2}+
\frac{e^\Im}{\sqrt{1+\hat\varpi^2}}\int\frac{\hat\varpi
e^{-\Im}}{\sqrt{(1+\hat\varpi^2)^3}}\:d\hat\varpi\right) +
\frac{\al+\be z^{-1}e^\Im}{\sqrt{1+\hat\varpi^2}}.\eqno(4.103)$$

 Modulo the transformation in (2.22) and (2.23), we take $g=\sgm^{-1}$.
According to (4.18), (4.27) and (4.101),
\begin{eqnarray*}\hspace{1cm}f_x&=&\sgm^{-1}(\gm_1-x^{-2}\phi)-\nu\sgm
=x^{-1}e^{-\Im-\gm_1}\left(\gm_1'
-\frac{\be_1}{x^2+z^2}\right)-\nu xe^{\Im+\gm_1}\\ & &+
\frac{x^{-3}e^{-\gm_1}}{\sqrt{1+\hat\varpi^2}}\left(\be+
\be_1\int\frac{\hat\varpi
e^{-\Im}}{\sqrt{(1+\hat\varpi^2)^3}}\:d\hat\varpi\right).\hspace{4.7cm}(4.104)\end{eqnarray*}
According (4.83), $\kappa_z=\tau_x$. Hence (4.16) gives
\begin{eqnarray*}\hspace{1cm}p&=&\frac{\rho}{2}(\be_1'\ln(x^2+z^2)-
\kappa^2-\tau^2- {\gm_1'}'(x^2+z^2))-\al'\rho\sqrt{x^2+z^2}\\
&
&+\rho\int\frac{\hat\varpi^{-1}e^\Im}{\sqrt{1+\hat\varpi^2}}\left(\be'+
\be_1'\int\frac{\hat\varpi
e^{-\Im}}{\sqrt{(1+\hat\varpi^2)^3}}\:d\hat\varpi\right)d\hat\varpi.
\hspace{3.5cm}(4.105)\end{eqnarray*}

 Recall $\xi=f+\kappa y+g
e^{\sgm y}$ and $\eta=\tau y+\zeta e^{\sgm y}$. By (4.1), (4.27),
(4.102) and (4.103), we have the following second theorem in this
section. \psp

{\bf Theorem 4.2}. {\it Let $\hat\varpi=x/z$. Suppose that
$\al,\be,\be_1,\gm_1$ are arbitrary functions of $t$ and $\Im$ is
any function of $\hat\varpi$. We have the following solutions of
the three-dimensional classical non-steady boundary layer
equations (1.3), (1.4) and (1.5):
\begin{eqnarray*}\hspace{2cm}u&=&x\gm_1'+\be_1x^{-1}\left(\frac{\hat\varpi^2}{1+\hat\varpi^2}-
\frac{e^\Im}{\sqrt{1+\hat\varpi^2}}\int\frac{\hat\varpi
e^{-\Im}}{\sqrt{(1+\hat\varpi^2)^3}}\:d\hat\varpi\right)
\\ & &+\frac{\al\hat\varpi-\be x^{-1}e^\Im}{\sqrt{1+\hat\varpi^2}}
+\exp\left(xye^{\Im+\gm_1}\right),\hspace{5.3cm}(4.106)\end{eqnarray*}
\begin{eqnarray*}\hspace{2cm}w&=&z\gm_1'+\be_1z^{-1}\left(\frac{1}{1+\hat\varpi^2}+
\frac{e^\Im}{\sqrt{1+\hat\varpi^2}}\int\frac{\hat\varpi
e^{-\Im}}{\sqrt{(1+\hat\varpi^2)^3}}\:d\hat\varpi\right)\\ & & +
\frac{\al+\be z^{-1}e^\Im}{\sqrt{1+\hat\varpi^2}}+
\frac{z}{x}\exp\left(xye^{\Im+\gm_1}\right),
\hspace{5.2cm}(4.107)\end{eqnarray*}
\begin{eqnarray*}v&=&\frac{e^\Im}{\sqrt{1+\hat\varpi^2}}\left(\frac{(x^2+z^2)\Im'y}{xz^3}-\frac{y}{x^2}-
\frac{e^{-\Im-\gm_1}}{x^3}\right) \left(\be+
\be_1\int\frac{\hat\varpi
e^{-\Im}}{\sqrt{(1+\hat\varpi^2)^3}}\:d\hat\varpi\right) +\nu
xe^{\Im+\gm_1}\\& &-\frac{\gm_1'e^{-\Im-\gm_1}}{x}+
\frac{\be_1(xy+e^{-\Im-\gm_1})}{x(x^2+z^2)}
-\left(2\gm_1'+\frac{\al}{\sqrt{x^2+z^2}}\right)y-
\frac{y}{x}\exp\left(xye^{\Im+\gm_1}\right)\hspace{0.3cm}(4.108)
\end{eqnarray*}
and $p$ is given in (4.105) with $\kappa$ in (4.102) and $\tau$ in
(4.103).}

\section{3-D Distinct Exponential Approach}

In this section, we will find certain function-parameter exact
solutions with two distinct exponential functions in $y$ for the
three-dimensional non-steady boundary layer equations (1.3)-(1.5).

We use the settings in (4.1)-(4.12) and continue our case-by-case
approach.\psp

{\it Case 2}.  $H=\Phi^{-1}=e^{\varpi}$. \psp

In this case,
\begin{eqnarray*}& &\xi_y\xi_{yx}-(\xi_x+\eta_z)\xi_{yy}+\eta_y\xi_{yz}
=\kappa\kappa_x+\tau\kappa_z-\sgm\zeta(\sgm g)_z-\sgm^2
 g\zeta_z-\sgm\zeta\kappa_ze^{-\varpi}+(\sgm g(\sgm
 g)_x\\ & &-\sgm^2 gg_x)e^{2\varpi}+
\{(\sgm g\kappa)_x+(\sgm g)_z\tau +\sgm g[(\kappa\sgm_x+\tau\sgm_z
-\sgm(\kappa_x+\tau_z))y-\sgm f_x]\}e^\varpi \hspace{1cm}(5.1)
\end{eqnarray*}
by (4.11) and
\begin{eqnarray*}& &\xi_y\eta_{yx}-(\xi_x+\eta_z)\eta_{yy}
+\eta_y\eta_{yz}\\ &=&\kappa\tau_x+\tau\tau_z-\sgm g(\sgm
\zeta)_x-\sgm^2 g_x\zeta+\sgm g\tau_xe^\varpi+(\sgm \zeta(\sgm
 \zeta)_z-\sgm^2 \zeta \zeta_z)e^{-2\varpi}\\ & &
+ \{-(\sgm \zeta\tau)_z-(\sgm \zeta)_x\kappa +\sgm
\zeta[(\kappa\sgm_x+\tau\sgm_z -\sgm(\kappa_x+\tau_z))y -\sgm
f_x]\}e^{-\varpi}\hspace{2.4cm}(5.2)
\end{eqnarray*}
by (4.12), where we treat $\varpi=\sgm y$. Moreover, (4.14) holds
and
$$\eta_{yt}=\tau_t+(\sgm\sgm_t\zeta y-(\sgm\zeta)_t)e^{-\varpi},
\qquad\eta_{yyy}=-\sgm^3\zeta e^{-\varpi}.\eqno(5.3)$$ Thus (4.2)
and (4.3) are implied by the following system of partial
differential equations:
$$\kappa_t+\kappa\kappa_x+\tau\kappa_z-\sgm\zeta(\sgm g)_z-\sgm^2
 g\zeta_z+\frac{1}{\rho}p_x=0,\eqno(5.4)$$
$$\tau_t+\kappa\tau_x+\tau\tau_z-\sgm g(\sgm
\zeta)_x-\sgm^2 g_x\zeta+\frac{1}{\rho}p_z=0,\eqno(5.5)$$
$$\kappa_z=0,\qquad\tau_x=0,\eqno(5.6)$$
$$\sgm_x=0,\qquad\sgm_z
 =0,\eqno(5.7)$$
$$\sgm_t+\kappa\sgm_x+\tau\sgm_z
-\sgm(\kappa_x+\tau_z)=0,\eqno(5.8)$$
$$(\sgm g)_t+(\sgm g\kappa)_x+(\sgm g)_z\tau -\sgm^2g(f_x+\nu\sgm)=0,
\eqno(5.9)$$
$$(\sgm\zeta)_t+(\sgm \zeta\tau)_z+(\sgm \zeta)_x\kappa +\sgm^2
\zeta(f_x-\nu\sgm)=0.\eqno(5.10)$$

According to (5.7),
$$\sgm=\gm(t)\eqno(5.11)$$
for some nonzero function $\gm$ of $t$.  Moreover, (5.8) can
written as
$$\kappa_x+\tau_z=\frac{\gm'}{\gm}.\eqno(5.12)$$
By (5.6) and (2.17)-(2.20), we take
$$\kappa=\frac{\gm'-\al'\gm}{\gm}x,\qquad\tau=\al'
z,\eqno(5.13)$$ where $\al$ is a function of $t$. Modulo the
transformation in (2.22) and (2.23), we take $g=\gm^{-1}$, and
(5.9) becomes
$$f_x=\frac{\gm'-\al'\gm}{\gm^2}-\nu\gm.\eqno(5.14)$$
Substituting (5.11) and (5.14) into (5.10), we get
$$2(\gm'-\nu\gm^3)\zeta+\gm\zeta_t+\al'\gm z\zeta_z+
(\gm'-\al'\gm)x\zeta_x =0.\eqno(5.15)$$ Thus
$$\zeta=\frac{e^{2\nu\int\gm^2dt}}{\gm^2}\phi(e^\al
x/\gm,e^{-\al}z).\eqno(5.16)$$

On the other hand, (5.4)-(5.6), (5.11) and (5.13) say that the
compatibility  $p_{xz}=p_{zx}$ in (5.4) and (5.5) is implied by
$$(\zeta)_{xx}=(\zeta)_{zz}.\eqno(5.17)$$
Let $\es=\pm 1$. The above two equations imply
$$\gm=\es e^{2\al},\qquad\zeta=\exp\left(2\nu\int
e^{4\al}dt-4\al\right)[\Im(e^{-\al}(x+z))+
\iota(e^{-\al}(x-z))],\eqno(5.18)$$ where $\Im$ and $\iota$ are
arbitrary one-variable functions.
 Moreover, (5.4) and
(5.5) yield \begin{eqnarray*}\hspace{2cm}p&=&\frac{\rho}{2}\{\es
\exp\left(2\nu\int e^{4\al}dt-2\al\right)[\Im(e^{-\al}(x+z))-
\iota(e^{-\al}(x-z))]
\\ & &-({\al'}'+(\al')^2)(x^2+z^2)\}\hspace{7.4cm}(5.19)\end{eqnarray*}
 modulo the
transformation in (2.21).

 By (4.1), we have the following
theorem.\psp

{\bf Theorem 5.1}. {\it Let $\Im,\iota$ be one-variable functions
and let $\al$ be function of $t$. Suppose $\es=\pm 1$. We have the
following solutions of the three-dimensional classical non-steady
boundary layer equations (1.3), (1.4) and (1.5):
$$u=\al'x+\exp(\es e^{2\al}y),\eqno(5.20)$$
$$w=\al'z-\es \exp\left(-\es e^{2\al}y+2\nu\int
e^{4\al}dt-2\al\right)[\Im(e^{-\al}(x+z))+
\iota(e^{-\al}(x-z))],\eqno(5.21)$$
\begin{eqnarray*}\hspace{2cm}v&=&\es(\nu e^{2\al}-\al' e^{2\al})-2\al' y
-\exp\left(-\es e^{2\al}y+2\nu\int
e^{4\al}dt-5\al\right)\\
& &\times[\Im'(e^{-\al}(x+z))-
\iota'(e^{-\al}(x-z))]\hspace{5.6cm}(5.22)\end{eqnarray*} and $p$
is given in (5.19).} \psp

{\it Case 3}. $H=e^\varpi$ and $\Phi=0=\zeta$. \psp

As in the earlier cases, we take $\sgm g=1$. Then
\begin{eqnarray*}\hspace{2cm}& &\xi_y\xi_{yx}-(\xi_x+\eta_z)\xi_{yy}+\eta_y\xi_{yz}
 =\kappa\kappa_x+\tau\kappa_z+\sgm_x\sgm^{-1}e^{2\varpi}\\ & &
+ \{\kappa_x+[(\kappa\sgm_x+\tau\sgm_z
-\sgm(\kappa_x+\tau_z))y-\sgm f_x]\}e^\varpi.\hspace{3.9cm}(5.23)
\end{eqnarray*}
 by (4.11) and
$$\xi_y\eta_{yx}-(\xi_x+\eta_z)\eta_{yy}
+\eta_y\eta_{yz}=\kappa\tau_x+\tau\tau_z+\sgm
g\tau_xe^\varpi.\eqno(5.24)$$ by (4.12).
 Thus (4.2) and (4.3) are
implied by the following system of partial differential equations:
(4.16),
$$\sgm_x=0=\tau_x,\qquad \sgm_t+\tau\sgm_z
-\sgm(\kappa_x+\tau_z)=0,\qquad\kappa_x=\sgm(f_x+\nu\sgm).\eqno(5.25)$$

We write
$$\sgm=e^{\vs(t,z)},\qquad\tau=\ves(t,z)\eqno(5.26)$$
for some functions $\vs$ and $\ves$ of $t$ and $z$. The second
equation in (5.25) yields
$$\kappa_x=\vs_t+\ves\vs_z-\ves_z=\psi(t,z)\;\;\mbox{is a function
of}\;\;t,z\lra \kappa=\phi(t,z)+\psi(t,z)x\eqno(5.27)$$ for some
function $\phi$ of $t$ and $z$. The compatibility $p_{xz}=p_{zx}$
in (4.16) is equivalent to
$$\ptl_z(\kappa_t+\kappa\kappa_x+\tau\kappa_z)=0.\eqno(5.28)$$
Note
$$\kappa\kappa_x+\tau\kappa_z=
\phi_t+\psi_tx+\psi\phi+\psi^2x+\ves\phi_z+\ves\psi_zx.\eqno(5.29)$$
For simplicity of solving (5.28) for $\phi,\psi$ and (5.27) for
$\vs$, we assume
$$\phi_t+\psi\phi+\ves\phi_z=0,\qquad
\psi_t+\psi^2+\ves\psi_z=0,\qquad\ves=-\frac{\al'(t)}{\iota'(z)}\eqno(5.30)$$
for some functions $\al$ of  $t$ and $\iota$ of $z$. Thus
$$\psi=\frac{1}{t+\Im(\al+\iota)},\qquad
\phi=\frac{\Im_1(\al+\iota)}{t+\Im(\al+\iota)},\qquad\vs=
\ln\Im_2(\al+\iota)(t+\Im(\al+\iota))-\ln\iota'\eqno(5.31)$$ for
some one-variable functions $\Im,\Im_1,\Im_2$. Hence
$$\kappa=\frac{x+\Im_1(\al+\iota)}{t+\Im(\al+\iota)},\qquad\tau=
-\frac{\al'}{\iota'},\eqno(5.32)$$
$$\sgm=\frac{\Im_2(\al+\iota)(t+\Im(\al+\iota))}{\iota'},\qquad g=
\frac{\iota'}{\Im_2(\al+\iota)(t+\Im(\al+\iota))},\eqno(5.33)$$
$$p=\frac{\rho}{2}\left(2{\al'}'\int\frac{dz}{\iota'}-
\left(\frac{\al'}{\iota'}\right)^2\right)\eqno(5.34)$$ modulo the
transformation in (2.21). Furthermore,  the last equation in
(5.25) yields
$$f_x=\frac{\iota'}{\Im_2(\al+\iota)(t+\Im(\al+\iota))^2}-
\frac{\nu}{\iota'}\Im_2(\al+\iota)(t+\Im(\al+\iota)).\eqno(5.35)$$
Recall $\xi=f+\kappa y+ge^{\sgm y}$ and $\eta=\tau y$. By (4.1),
we obtain the following theorem:\psp

{\bf Theorem 5.2}. {\it Let $\iota$ be any function of $z$ and let
$\al$ be  any function of $t$. Suppose that $\Im,\;\Im_1$ and
$\Im_2$ are arbitrary one-variable functions. We have the
following solutions of the three-dimensional classical non-steady
boundary layer equations (1.3), (1.4) and (1.5):
$$u=\frac{x+\Im_1(\al+\iota)}{t+\Im(\al+\iota)}+\exp[y(\iota')^{-1}
\Im_2(\al+\iota)(t+\Im(\al+\iota))],\qquad
w=-\frac{\al'}{\iota'},\eqno(5.36)$$
$$v=\frac{\nu}{\iota'}\Im_2(\al+\iota)(t+\Im(\al+\iota))-
\frac{\iota'}{\Im_2(\al+\iota)(t+\Im(\al+\iota))^2}-
\frac{y}{t+\Im(\al+\iota)}\eqno(5.37)$$ and $p$ is given in
(5.34).}\psp

{\it Case 4}. $H=e^\varpi$ and $\Phi=e^{\gm\varpi}$ for a function
$\gm$ of $t$.\psp

Again we assume $\sgm g=1$. In this case,
\begin{eqnarray*}& &\xi_y\xi_{yx}-(\xi_x+\eta_z)\xi_{yy}+\eta_y\xi_{yz}
=\kappa\kappa_x+\tau\kappa_z+\gm\sgm\zeta\kappa_ze^{\gm\varpi}
 -\sgm\zeta_ze^{(1+\gm)\varpi}
\\ & &+
\{\kappa_x+[(\kappa\sgm_x+\tau\sgm_z -\sgm(\kappa_x+\tau_z))y
-\sgm f_x]\}e^{\varpi} +\sgm_x ge^{2\varpi}.\hspace{4cm}(5.38)
\end{eqnarray*} by (4.11) and
\begin{eqnarray*}& &\xi_y\eta_{yx}-(\xi_x+\eta_z)\eta_{yy}
+\eta_y\eta_{yz}=\kappa\tau_x+\tau\tau_z+\sgm g\tau_xe^\varpi
\\ & &+
\gm\{(\sgm \zeta\tau)_z+(\sgm \zeta)_x\kappa+\gm\sgm
\zeta[(\kappa\sgm_x+\tau\sgm_z -\sgm(\kappa_x+\tau_z))y -\sgm
f_x]\}e^{\gm\varpi}
\\ & &+\gm^2\sgm\sgm_z\zeta^2e^{2\gm\varpi}
+\gm\sgm (g(\sgm \zeta)_x-\gm\sgm
 g_x\zeta)e^{(1+\gm)\varpi}.\hspace{6.1cm}(5.39)
\end{eqnarray*}
by (4.12). Moreover, we have (4.14) and
$$\eta_{yt}=\tau_t+((
\gm\sgm\zeta)_t+\gm\sgm\sgm_t\zeta y)e^{\gm\varpi},\qquad
\eta_{yyy}=(\gm\sgm)^3\zeta e^{\gm\varpi}.\eqno(5.40)$$ Thus (4.2)
and (4.3) are implied by the following system of partial
differential equations: (4.16),
$$\kappa_z=\zeta_x=\zeta_z=\sgm_z=\sgm_x=\tau_x=0,\eqno(5.41)$$
$$\sgm_t-\sgm(\kappa_x+\tau_z)=0,\qquad\kappa_x=\sgm(f_x+\nu\sgm)
\eqno(5.42)$$
$$(\gm\sgm\zeta)_t+\gm\sgm \zeta\tau_z=
\gm^2\sgm^2 \zeta (f_x+\nu\gm\sgm).\eqno(5.43)$$

According to (5.41) and (5.42),
$$\sgm=\be,\qquad\kappa=\al'x,\qquad\tau=
\frac{(\be'-\al'\be)z}{\be} \eqno(5.44)$$ modulo the
transformations in (2.17)-(2.20), where $\al$ and $\be$ are
arbitrary function of $t$. Moreover,
$$f_x=\frac{\al'}{\be}-\nu\be.\eqno(5.45)$$
Then (5.43) becomes
$$(\ln\gm\be\zeta)_t+\frac{\be'-\al'\be}{\be}=
\gm(\al'+\nu(\gm-1)\be^2).\eqno(5.46)$$ So
$$\zeta=\frac{1}{\gm\be^2}e^{\al+\int\gm(\al'+\nu(\gm-1)\be^2)dt}.
\eqno(5.47)$$ Now
$$p=-\frac{\rho}{2}[({\al'}'+(\al')^2)x^2
+({\be'}'-2\al'\be'+((\al')^2-{\al'}')\be)\be^{-1}z^2]\eqno(5.48)$$
 modulo the transformation in (2.21). Moreover,
$$\xi=f+\al'xy+\frac{1}{\be}e^{\be y},\eqno(5.49)$$
$$\eta=\frac{(\be'-\al'\be)yz}{\be}+
\frac{1}{\gm\be^2}e^{\al+\be\gm
y+\int\gm(\al'+\nu(\gm-1)\be^2)dt}. \eqno(5.50)$$ By (4.1), we
have the following theorem.\psp

{\bf Theorem 5.3}. {\it Let $\al,\be$ and $\gm$ be any functions
of $t$. We have the following solutions of the three-dimensional
classical non-steady boundary layer equations (1.3), (1.4) and
(1.5):
$$u=\al'x+e^{\be y},\qquad
v=\nu\be-\frac{\al'}{\be}-\frac{\be'}{\be}y,\eqno(5.51)$$
$$w=\frac{(\be'-\al'\be)z}{\be} +\frac{1}{\be}e^{\al+\be\gm
y+\int\gm(\al'+\nu(\gm-1)\be^2)dt}\eqno(5.52)$$ and $p$ is given
in (5.48).}

\section{3-D Trigonometric and Hyperbolic Approach}

In this section, we will find certain function-parameter exact
solutions with trigonometric and hyperbolic-type functions in $y$
for the three-dimensional non-steady boundary layer equations
(1.3)-(1.5).

We start with the general approach given in (4.1)-(4.12) and
continue the case-by-case process. Again we use the notion
$\varpi=\sgm(t,x,z)y$. For $a\in\mbb{R}$, we denote
$$\vt_0=\frac{e^{\varpi}-ae^{-\varpi}}{2},\qquad\hat\vt_0=
\frac{e^{\varpi}+ae^{-\varpi}}{2},\eqno(6.1)$$
$$\vt_1=\sin\varpi,\qquad \hat\vt_1=\cos\varpi.\eqno(6.2)$$
\pse

{\it Case 5}. $H=\vt_r$ with $r=0,1,$ and $\zeta=0=\Phi$.\psp

 Note
\begin{eqnarray*}& &\xi_y\xi_{yx}-(\xi_x+\eta_z)\xi_{yy}+\eta_y\xi_{yz}
 =\kappa\kappa_x+\tau\kappa_z+(a\dlt_{0,r}+\dlt_{1,r})\sgm^2 gg_x+[(\sgm g\kappa)_x+(\sgm
g)_z\tau]\hat\vt_r\\
& &-(-1)^r\sgm g\{[\kappa\sgm_x+\tau\sgm_z
-\sgm(\kappa_x+\tau_z)]y-\sgm f_x\}\vt_r+\sgm\sgm_x
g^2\hat\vt_r^2\hspace{4cm}(6.3)
\end{eqnarray*}
by (4.11) and $$\xi_y\eta_{yx}-(\xi_x+\eta_z)\eta_{yy}
+\eta_y\eta_{yz}=\kappa\tau_x+\tau\tau_z+\sgm
g\tau_x\hat\vt_r\eqno(6.4)$$ by (4.12).
 Moreover,
$$\xi_{yt}=\kappa_t+(\sgm g)_t
\hat\vt_r-(-1)^r\sgm\sgm_tgy\vt_r,\;\;
\xi_{yyy}=(-1)^r\sgm^3g\hat\vt_r,\;\;\eta_{yt}=\tau_t,\;\;
\eta_{yyy}=0\eqno(6.5)$$ by (4.4) and (4.5). Thus (4.2) and (4.3)
are implied by the following system of partial differential
equations:
$$\tau_x=\sgm_x=f_x=0,\eqno(6.6)$$
$$\kappa_t+\kappa\kappa_x+\tau\kappa_z+(a\dlt_{0,r}+\dlt_{1,r})\sgm^2
gg_x+\frac{1}{\rho}p_x=0,\qquad\tau_t
+\tau\tau_z+\frac{1}{\rho}p_z=0,\eqno(6.7)$$
$$(\sgm g)_t+(\sgm g\kappa)_x+(\sgm
g)_z\tau-(-1)^r\nu\sgm^3g=0,\eqno(6.8)$$
$$\sgm_t+\tau\sgm_z -\sgm(\kappa_x+\tau_z)=0.\eqno(6.9)$$

We take $f=0$ by (6.6). Let $\al$ be a function of $t$ and let
$\iota$ be a function of $z$. Denote
$$\hat\varpi=\al-\iota,\qquad\td\varpi=\frac{x}{\sqrt{\al'}}.\eqno(6.10)$$
In order to solve (6.9), we first assume
$$\tau=\frac{\al'}{\iota'},\qquad
\kappa=\frac{{\al'}'}{2\al'}x.\eqno(6.11)$$ Then
$$\sgm=\frac{\sqrt{\al'}\Im(\hat\varpi)}{\iota'}\eqno(6.12)$$
for some one-variable function $\Im$. Multiplying (6.8) by $2\sgm
g$, we obtain
$$[(\sgm g)^2]_t+\kappa[(\sgm g)^2]_x+\tau[(\sgm
g)^2]_z+2(\kappa_x-(-1)^r\nu\sgm^2)(\sgm g)^2=0.\eqno(6.13)$$ So
$$(\sgm
g)^2=\frac{1}{\al'}\vs(\hat\varpi,\td\varpi)
\exp\left((-1)^r2\nu\Im^2\int\frac{dz}{(\iota')^2}\right)\eqno(6.14)$$
for some two-variable function $\vs$. By the compatibility
$p_{xz}=p_{zx}$, $\ptl_x\ptl_z(\sgm g)^2=0$. Hence we take
$$g=\frac{\iota'\vf(\hat\varpi)}{\al'\Im(\hat\varpi)}
\exp\left((-1)^r2\nu(\Im(\hat\varpi))^2\int\frac{dz}{(\iota')^2}\right),\eqno(6.13)$$
where $\vf$ is a one-variable function. According to (6.5) and
(6.9),
$$p=-\frac{\rho}{2}\left(\frac{2\al'{{\al'}'}'-({\al'}')^2}
{4(\al')^2}x^2+\frac{(\al')^2}
{(\iota')^2}+{\al'}'\int\frac{dz}{\iota'}\right)\eqno(6.14)$$
modulo the transformation in (2.21).

Recall $\xi=\kappa y+g H(\varpi)$ and $\eta=\tau y$.
 By (4.1), we have the
following theorem. \psp

{\bf Theorem 6.1}. {\it Let $\iota$ be any function of $z$ and let
$\al$ be an arbitrary function of $t$.
 Suppose that $\Im,\vf$ are any functions of $\hat\varpi=\al-\iota$ and
 $a\in\mbb{R}$.
We have the following solutions of the three-dimensional classical
non-steady boundary layer equations (1.3), (1.4) and (1.5):
$$w=\frac{\al'}{\iota'},\qquad v=
\frac{\al'{\iota'}'y}{(\iota')^2}-\frac{{\al'}'}{2\al'}y,
\eqno(6.15)$$ $p$ is given in (6.14), and
\begin{eqnarray*}\hspace{2cm}u&=&\frac{{\al'}'}{2\al'}x+\frac{\vf}{2\sqrt{\al'}}
\left[\exp\left(2\nu\Im^2\int\frac{dz}{(\iota')^2}\right)
\right]\\ & &\times \left[\exp\left(\frac{\sqrt{\al'}\Im
y}{\iota'}\right)+a\exp\left(-\frac{\sqrt{\al'}\Im
y}{\iota'}\right)\right]\hspace{4.2cm}(6.16)\end{eqnarray*} or
$$u=\frac{{\al'}'}{2\al'}x-\frac{\vf}{\sqrt{\al'}}
\left[\exp\left(-2\nu\Im^2\int\frac{dz}{(\iota')^2}\right)
\right]\cos\left(\frac{\sqrt{\al'}\Im
y}{\iota'}\right).\eqno(6.17)$$}\pse

Suppose
$$\kappa=\al'(t)x,\qquad \tau=\be'(t)z,\qquad \sgm=e^{\al+\be}\eqno(6.18)$$
for some functions $\al$ and $\be$ of $t$. Equation (6.9)
naturally holds. By the compatibility $p_{xz}=p_{zx}$ in (6.7), we
take
$$(\sgm g)^2=\ves(t,z)-\vs(t,x)\eqno(6.19)$$
for some two-variable functions $\ves$ and $\vs$. Now (6.13) is
implied by the system of the following equations:
$$\ves_t+\be'z\ves_z+2(\al'-(-1)^r\nu e^{2(\al+\be)})\ves=0,\qquad
\vs_t+\al'x\vs_x+2(\al'-(-1)^r\nu
e^{2(\al+\be)})\vs=0.\eqno(6.20)$$ Thus
$$\ves=\Im(ze^{-\be})\exp\left(-2\al+(-1)^r2\nu\int e^{2(\al+\be)}dt\right)
\eqno(6.21)$$ and
$$\vs=\iota(xe^{-\al})\exp\left(-2\al+(-1)^r2\nu\int e^{2(\al+\be)}dt\right),
\eqno(6.22)$$ where $\iota$ and $\Im$ are arbitrary one-variable
functions. Hence
$$g=\sqrt{\Im(ze^{-\be})-\iota(xe^{-\al})}\exp\left(-\be-2\al
+(-1)^r\nu\int e^{2(\al+\be)}dt\right)\eqno(6.23)$$ and
\begin{eqnarray*}\hspace{2cm}p&=&\frac{\rho}{2}[(a\dlt_{0,r}+\dlt_{1,r})\iota(xe^{-\al})\exp\left((-1)^r2\nu\int
e^{2(\al+\be)}dt-2\al\right)\\ & &-({\al'}'+(\al')^2)x^2
-({\be'}'+(\be')^2)z^2]\hspace{5.7cm}(6.24)\end{eqnarray*}
 modulo the transformation in (2.21).

Recall $\xi=\kappa y+g H(\varpi)$ and $\eta=\tau y$.
 By (4.1), we have the
following theorem. \psp

{\bf Theorem 6.2}. {\it Let $\iota,\Im$ be any one-variable
functions and let $\al,\be$ be any functions of $t$. We have the
following solutions of the three-dimensional classical non-steady
boundary layer equations (1.3), (1.4) and (1.5): (1) $w=\be'z$,
$$u=\al' x+\sqrt{\Im(ze^{-\be})-\iota(xe^{-\al})}
\exp\left(-\al-\nu\int e^{2(\al+\be)}dt\right)\;\cos
(e^{\al+\be}y),\eqno(6.25)$$
\begin{eqnarray*}\hspace{1.2cm}v&=&\frac{e^{-\al}\iota'(xe^{-\al})}{2
\sqrt{\Im(ze^{-\be})-\iota(xe^{-\al})}}
\exp\left(-\be-2\al-\nu\int e^{2(\al+\be)}dt\right)\\ &
&\times\sin (e^{\al+\be}y)
-(\al'+\be')y\hspace{8cm}(6.26)\end{eqnarray*} and $p$ is given in
(6.24) with $r=1$; (2) $w=\be'z$,
\begin{eqnarray*}\hspace{1.2cm}u&=&\al'
x+\frac{1}{2}\sqrt{\Im(ze^{-\be})-\iota(xe^{-\al})}
\left[\exp\left(-\al+\nu\int e^{2(\al+\be)}dt\right)\right]\\ &
&\times \left[ \exp(e^{\al+\be}y)+a\exp(-e^{\al+\be}y)\right]
\hspace{6.6cm}(6.27)\end{eqnarray*}
\begin{eqnarray*}\hspace{1cm}v&=&\frac{e^{-\al}\iota'(xe^{-\al})}{4
\sqrt{\Im(ze^{-\be})-\iota(xe^{-\al})}}
\left[\exp\left(-\be-2\al+\nu\int e^{2(\al+\be)}dt\right)\right]\\
& &\times \left[ \exp(e^{\al+\be}y)-a\exp(-e^{\al+\be}y)\right]
-(\al'+\be')y\hspace{4.6cm}(6.28)\end{eqnarray*} for
$a\in\mbb{R}$, and $p$ is given in (6.24) with $r=0$.} \psp

{\it Case 6}. $H=\Phi=\vt_r$ in (6.1) and (6.2), and
$\sgm=e^{\al(t)}$ for some function $\al$ of $t$.\psp

Note $\varpi=e^\al y$ in this case. Moreover,
\begin{eqnarray*}& &\xi_y\xi_{yx}-(\xi_x+\eta_z)\xi_{yy}+\eta_y\xi_{yz}
=\kappa\kappa_x+\tau\kappa_z -(-1)^re^{2\al} g[(\kappa_x+\tau_z)y+
f_x]\vt_r\\
& &+e^\al[(g\kappa)_x+g_z\tau+\zeta\kappa_z]
\hat\vt_r+(a\dlt_{0,r}+\dlt_{1,r})e^{2\al}g(g_x+\zeta_z)+e^{2\al}(\zeta
g_z- g\zeta_z) \hat\vt_r^2\hspace{1.2cm}(6.29)
\end{eqnarray*}
by (4.11) and
$$\xi_{yt}=\kappa_t+(\al'g+g_t)e^\al
\hat\vt_r+(-1)^r\al'e^{2\al}gy\vt_r,\qquad
\xi_{yyy}=(-1)^re^{3\al} g\hat\vt_r.\eqno(6.30)$$

By symmetry, (4.2) and (4.3) are implied by the following system
of partial differential equations: $f_x=0$,
$$\al'-\kappa_x-\tau_z=0,\eqno(6.31)$$
$$\zeta g_z-
 g\zeta_z=0,\qquad g\zeta_x-g_x\zeta=0
.\eqno(6.32)$$
$$\kappa_t+\kappa\kappa_x+\tau\kappa_z+(a\dlt_{0,r}+\dlt_{1,r})e^{2\al}g(g_x+\zeta_z)
+\frac{1}{\rho}p_x=0,\eqno(6.33)$$
$$\tau_t+\kappa\tau_x+\tau\tau_z+(a\dlt_{0,r}+\dlt_{1,r})e^{2\al}\zeta(g_x+\zeta_z)
+\frac{1}{\rho}p_z=0,\eqno(6.34)$$
$$\al'g+ g_t+(g\kappa)_x+g_z\tau+\zeta\kappa_z-(-1)^r\nu
e^{2\al}g=0,\eqno(6.35)$$
$$\al'\zeta+ \zeta_t+(\zeta\tau)_z+ \zeta_x\kappa+
g\tau_x-(-1)^r\nu e^{2\al}\zeta=0.\eqno(6.36)$$

We take $f=0$ by (4.1). Moreover, (6.31) yields
$$\kappa=\al'x-\vs_z,\qquad \tau=\vs_x\eqno(6.37)$$
for some function $\vs$ of $t,x,z$. By (6.32),
$$\zeta=g\tan\gm\eqno(6.38)$$
for some function $\gm$ of $t$. According to (4.28) and (4.29),
(6.35) and (6.36) give
$$\gm'\sec^2\gm+\tau_x
-\kappa_z\tan^2\gm+ (\tau_z-\kappa_x)\tan\gm=0,\eqno(6.39)$$
equivalently,
$$\gm'\sec^2\gm+\vs_{xx}
+\vs_{zz}\tan^2\gm+ 2\vs_{xz}\tan\gm-\al'\tan\gm=0.\eqno(6.40)$$
Now we use the notations in (4.31). The above equation is
equivalent to
$$\ptl_{\td\varpi}^2(\vs)=\al'\sin\gm\;\cos\gm-\gm',\eqno(6.41)$$
that is,
$$\vs=\phi(t,\hat\varpi)+\psi(t,\hat\varpi)\td\varpi+\frac{1}{2}(\al'\sin\gm\;
\cos\gm-\gm')\td\varpi^2\eqno(6.42)$$ for some functions $\phi$
and $\psi$ of $t$ and $\hat\varpi$. So
$$\kappa=\al'x-(\phi_{\hat\varpi}+\td\varpi\psi_{\hat\varpi})\cos\gm-
\psi\sin\gm-(\al'\sin\gm\;
\cos\gm-\gm')\td\varpi\sin\gm,\eqno(6.43)$$
$$\tau=-(\phi_{\hat\varpi}+\td\varpi\psi_{\hat\varpi})\sin\gm+
\psi\cos\gm+(\al'\sin\gm\;
\cos\gm-\gm')\td\varpi\cos\gm.\eqno(6.44)$$

Note
$$\tau_x-\kappa_z=\phi_{\hat\varpi\hat\varpi}+\td\varpi
\psi_{\hat\varpi\hat\varpi}+\al'\sin\gm\;\cos\gm-\gm'.\eqno(6.45)$$
For simplicity of solving the problem, we assume
$$\phi=-\frac{1}{2}(\al'\sin\gm\;\cos\gm-\gm')\hat\varpi^2,\qquad
\psi_{\hat\varpi\hat\varpi}=0,\eqno(6.46)$$ that is,
$\tau_x-\kappa_z=0$.  The compatibility $p_{xz}=p_{zx}$ in (6.33)
and (6.36) is equivalent to
$\ptl_{\hat\varpi}\ptl_{\td\varpi}(g^2)=0$. So
$$g^2=\ves(t,\hat\varpi)+\vs(t,\td\varpi)\eqno(6.47)$$
for some function $\ves$ of $t$ and $\hat\varpi$, and some
function $\vs$ of $t$ and $\td\varpi$. On the other hand,  (6.35)
can be written as
$$ g_t+\kappa g_x+\tau g_z+
(\al'+\kappa_x+\kappa_z\tan\gm-(-1)^r\nu
e^{2\al})g=0,\eqno(6.48)$$ equivalently,
\begin{eqnarray*}\hspace{1cm} & &g_t+\al'x\ptl_x(g)-(\phi_{\hat\varpi}+
\td\varpi\psi_{\hat\varpi})\ptl_{\td\varpi}(g)
+(\psi+(\al'\sin\gm\;\cos\gm-\gm')\td\varpi)\ptl_{\hat\varpi}(g)\\
& &+(\al'(2-\sin^2\gm)-\psi_{\hat\varpi}+\gm'\tan\gm -(-1)^r\nu
e^{2\al})g=0.\hspace{3.9cm}(6.49)\end{eqnarray*} By (4.31) and
(4.32), we obtain
\begin{eqnarray*}\hspace{1cm}&&g_t+[\td\varpi(\al'\cos^2\gm-\psi_{\hat\varpi})
+\gm'\hat\varpi]\ptl_{\td\varpi}(g)
+(\al'\hat\varpi\sin^2\gm-\gm'\td\varpi+\psi)\ptl_{\hat\varpi}(g)\\
& &+(\al'(2-\sin^2\gm)-\psi_{\hat\varpi}+\gm'\tan\gm-(-1)^r\nu
e^{2\al})g=0.\hspace{3.9cm}(6.50)\end{eqnarray*}
 Multiplying $2g$ to
the above equation, we have
\begin{eqnarray*}\hspace{1cm}&&(g^2)_t+[\td\varpi(\al'\cos^2\gm-\psi_{\hat\varpi})
+\gm'\hat\varpi](g^2)_{\td\varpi}
+(\al'\hat\varpi\sin^2\gm-\gm'\td\varpi+\psi)(g^2)_{\hat\varpi}\\
& &+2(\al'(2-\sin^2\gm)-\psi_{\hat\varpi}+\gm'\tan\gm -(-1)^r\nu
e^{2\al})g^2=0.\hspace{3.5cm}(6.51)\end{eqnarray*} Substituting
(6.47) into the above equation, we get
\begin{eqnarray*}&&\ves_t+\vs_t+[\td\varpi(\al'\cos^2\gm-\psi_{\hat\varpi})
+2\gm'\hat\varpi]\vs_{\td\varpi}
+(\al'\hat\varpi\sin^2\gm-2\gm'\td\varpi+\psi)\ves_{\hat\varpi}\\
& &+2(\al'(2-\sin^2\gm)-\psi_{\hat\varpi}+\gm'\tan\gm-(-1)^r\nu
e^{2\al})(\ves+\vs)=0.\hspace{3.7cm}(6.52)\end{eqnarray*}

In order to solve (6.52), we assume
$$\ves_{\hat\varpi}=2\be\hat\varpi+\be_1,\qquad
\vs_{\td\varpi}=2\be\td\varpi+\be_2\eqno(6.53)$$ for some
functions $\be,\;\be_1,\;\be_2$ of $t$. So $\ves$ is a quadratic
polynomial in $\hat\varpi$  and $\vs$ is a quadratic polynomial in
$\td\varpi$, whose coefficients are functions of $t$. Now (6.52)
is implied by the following system of partial differential
equations:
\newpage
\begin{eqnarray*}\hspace{1.3cm}&&\ves_t+2\be_2\gm'\hat\varpi+
(\al'\hat\varpi\sin^2\gm+\psi)(2\be\hat\varpi+\be_1)\\
&&+2(\al'(2-\sin^2\gm)-\psi_{\hat\varpi}+\gm'\tan\gm -(-1)^r\nu
e^{2\al})\ves=0, \hspace{3.3cm}(6.54)\end{eqnarray*}
\begin{eqnarray*}\hspace{1.3cm}&&\vs_t-2\be_1\gm'\td\varpi+
[\td\varpi(\al'\cos^2\gm-\psi_{\hat\varpi})
](2\be\td\varpi+\be_2)\\
&&+2(\al'(2-\sin^2\gm)-\psi_{\hat\varpi}+\gm'\tan\gm -(-1)^r\nu
e^{2\al})\vs=0. \hspace{3.4cm}(6.55)\end{eqnarray*} By the
coefficients of quadratic terms, we have
$$\be'+2(2\al'+\gm'\tan\gm -(-1)^r\nu
e^{2\al})\be=0,\eqno(6.56)$$
$$\be'+2(\al'(2+\cos2\gm)-2\psi_{\hat\varpi}+\gm'\tan\gm -(-1)^r\nu
e^{2\al})\be=0,\eqno(6.57)$$ which implies
$$\psi_{\hat\varpi}=\frac{\al'}{2}\cos 2\gm.\eqno(6.58)$$ For
simplicity, we take
$$\psi=\frac{\al'}{2}\hat\varpi\cos 2\gm.\eqno(6.59)$$
According (6.56),
$$\be=b_1\left[\exp\left((-1)^r2\nu\int
e^{2\al}dt-4\al\right)\right]\cos^2\gm\eqno(6.60)$$ for
$b_1\in\mbb{R}$.

The coefficients of $\hat\varpi$ in (6.54) and of $\td\varpi$ in
(6.55) give
$$\be_1'+2\be_2\gm+
(7\al'/2+2\gm'\tan\gm-(-1)^r2\nu e^{2\al})\be_1=0,\eqno(6.61)$$
$$\be_2'-2\be_1\gm'+(7\al'/2+2\gm'\tan\gm +2\nu
e^{2\al})\be_2=0.\eqno(6.62)$$ So
$$\be_1=(b_2\cos 2\gm\;+b_3\sin 2\gm)\left[\exp\left((-1)^r2\nu\int
e^{2\al}dt-7\al/2\right)\right]\cos^2\gm,\eqno(6.63)$$
$$\be_2=(-b_2\sin 2\gm\;+b_3\cos
2\gm)\left[\exp\left((-1)^r2\nu\int
e^{2\al}dt-7\al/2\right)\right]\cos^2\gm\eqno(6.64)$$
 for $b_2,b_3\in\mbb{R}$.
According to (6.47), we take the zero constant term of $\vs$ and
the constant term of $\ves$ to be
$$b_4\left[\exp\left((-1)^r2\nu\int
e^{2\al}dt-3\al\right)\right]\cos^2\gm\eqno(6.65)$$
 by (6.54) with $b_4\in\mbb{R}$. Hence we take
\begin{eqnarray*}g&=&\sqrt{b_1e^{-\al}(\hat\varpi^2+\td\varpi^2)+e^{-\al/2}[
(b_2\hat\varpi+b_3\td\varpi)\cos
2\gm+(b_3\hat\varpi-b_2\td\varpi)\sin 2\gm]+b_4}\\ &
&\times\left[\exp\left((-1)^r\nu\int
e^{2\al}dt-3\al/2\right)\right]\cos\gm.
\hspace{6.5cm}(6.66)\end{eqnarray*} Furthermore,
$$\kappa=\frac{\al'x}{2}-\gm'(z\cos 2\gm-x\sin
2\gm) ,\eqno(6.67)$$
$$\tau=\frac{\al'z}{2}-\gm'(x\cos 2\gm+z\sin 2\gm)\eqno(6.68)$$
by (4.31), (6.43), (6.44), (6.46) and (6.58).

Set
$$\hat\kappa=\frac{\al'}{4}(x^2+z^2)+\frac{\gm'\sin 2\gm}{2}(x^2-z^2)-xz\gm'\cos 2\gm
.\eqno(6.69)$$ Then
$$\kappa=\hat\kappa_x,\;\;\tau=\hat\kappa_z\lra\kappa_t=(\hat\kappa_t)_x,\;\;
\tau_t=(\hat\kappa_t)_z.\eqno(6.70)$$
 Moreover, \begin{eqnarray*}  & &e^{2\al}g(g_x+\zeta_z)=
\frac{\sec\gm}{2}\ptl_{\td\varpi}(e^{2\al}g^2)\\
&=&\frac{\cos\gm}{2} [2b_1e^{-\al}\td\varpi+e^{-\al/2}(b_3\cos
2\gm-b_2\sin 2\gm)] \left[\exp\left((-1)^r2\nu\int
e^{2\al}dt-\al\right)\right] . \hspace{0.4cm}(6.71)\end{eqnarray*}
By  (6.33), (6.34) and (6.69)-(6.71),
 \begin{eqnarray*}p&=&\rho\{xz({\gm'}'\cos2\gm+\al'\gm'\cos2\gm-2(\gm')^2\sin2\gm)+
 \frac{{\gm'}'\sin 2\gm+2(\gm')^2\cos\gm}{2}(z^2-x^2)\\ &&-
 \frac{4(\gm')^2+2{\al'}'+(\al')^2}{8}(x^2+z^2)-\frac{a\dlt_{0,r}+\dlt_{1,r}}{2}
\left[\exp\left((-1)^r2\nu\int e^{2\al}dt-\al\right)\right]
\\ & &\times[b_1e^{-\al}\td\varpi^2+e^{-\al/2}(b_3\cos
2\gm-b_2\sin 2\gm)\td\varpi]\} \hspace{5.9cm}(6.72)\end{eqnarray*}
 modulo the transformation in (2.21).

  Recall $\xi=\kappa y+g H(\varpi)$ and $\eta=\tau
y+\zeta \Phi$. Moreover, $\hat\varpi^2+\td\varpi^2=x^2+z^2$
according to (4.31).
 By (4.1), we have the
following theorem. \psp

{\bf Theorem 6.3}. {\it Let $\al,\gm$ be any functions of $t$ and
let $a,b_1,b_2,b_3,b_4\in\mbb{R}$. For $r=0,1$, we define $\vt_r$
and $\hat\vt_r$ in (6.1) and (6.2) with $\varpi=e^\al y$, and
$\hat\varpi$ and $\td\varpi$ in (4.31). We have the following
solutions of the three-dimensional classical non-steady boundary
layer equations (1.3), (1.4) and (1.5):
\begin{eqnarray*}u&=&\frac{\al'x}{2}+\sqrt{b_1e^{-\al}(x^2+z^2)+e^{-\al/2}[
(b_2\hat\varpi+b_3\td\varpi)\cos
2\gm+(b_3\hat\varpi-b_2\td\varpi)\sin 2\gm]+b_4}\\ &
&\times\left[\exp\left((-1)^r\nu\int
e^{2\al}dt-\al/2\right)\right]\hat\vt_r\cos\gm-\gm'(z\cos
2\gm-x\sin 2\gm),\hspace{1.6cm}(6.73)\end{eqnarray*}
\begin{eqnarray*}w&=&\frac{\al'z}{2}+\sqrt{b_1e^{-\al}(x^2+z^2)+e^{-\al/2}[
(b_2\hat\varpi+b_3\td\varpi)\cos
2\gm+(b_3\hat\varpi-b_2\td\varpi)\sin 2\gm]+b_4}\\ &
&\times\left[\exp\left((-1)^r\nu\int
e^{2\al}dt-\al/2\right)\right]\hat\vt_r\sin\gm-\gm'(x\cos
2\gm+z\sin 2\gm),\hspace{1.6cm}(6.74)\end{eqnarray*}
\begin{eqnarray*}v&=&-\frac{2b_1e^{-\al}\td\varpi+e^{-\al/2}(
b_3\cos 2\gm-b_2\sin
2\gm)}{2\sqrt{b_1e^{-\al}(x^2+z^2)+e^{-\al/2}[
(b_2\hat\varpi+b_3\td\varpi)\cos
2\gm+(b_3\hat\varpi-b_2\td\varpi)\sin 2\gm]+b_4}}\\ & &\times\\ &
&\times\left[\exp\left((-1)^r\nu\int
e^{2\al}dt-3\al/2\right)\right]\vt_r-\al'
y\hspace{5.8cm}(6.75)\end{eqnarray*}
 and $p$ is given in (6.72).}
\psp

{\it Case 7}. $H=\vt_r$ and $\Phi=\hat\vt_r$ in (6.1) and (6.2)
with $r=0,1$.\psp

In this case,
\begin{eqnarray*}& &\xi_y\xi_{yx}-(\xi_x+\eta_z)\xi_{yy}+\eta_y\xi_{yz}
\\ &=&\kappa\kappa_x+\tau\kappa_z+(a\dlt_{0,r}+\dlt_{1,r})\sgm^2 gg_x+[(\sgm g\kappa)_x+ (\sgm
g)_z\tau]\hat\vt_r+\sgm \sgm_xg^2\hat\vt_r^2
+(-1)^r\sgm\{\zeta\kappa_z\\ & &+g[(\kappa\sgm_x+\tau\sgm_z
-\sgm(\kappa_x+\tau_z))y -\sgm f_x]\}\vt_r +(-1)^r
 [\sgm\zeta(\sgm g)_z-\sgm^2
 g\zeta_z]\vt_r\hat\vt_r\hspace{1.1cm}(6.76)
\end{eqnarray*}
by (4.11), and
\begin{eqnarray*}& &\xi_y\eta_{yx}-(\xi_x+\eta_z)\eta_{yy}
+\eta_y\eta_{yz}\\
&=&\kappa\tau_x+\tau\tau_z+(-1)^r\{[(\sgm \zeta\tau)_z+(\sgm
\zeta)_x\kappa]\vt_r-(a\dlt_{0,r}+\dlt_{1,r})\sgm^2 \zeta
\zeta_z\}+\sgm\sgm_z \zeta^2\vt_r^2+ \sgm\{ g\tau_x\\ &
&+(-1)^r\zeta[(\kappa\sgm_x+\tau\sgm_z
-\sgm(\kappa_x+\tau_z))y-\sgm f_x]+(-1)^r[g(\sgm \zeta)_x-\sgm
g_x\zeta ]\vt_r\}\hat\vt_r\hspace{0.9cm}(6.77)
\end{eqnarray*}
by (4.12). Moreover, $$\xi_{yt}=\kappa_t+(\sgm g)_t
\hat\vt_r+(-1)^r\sgm\sgm_tgy\vt_r,\;\;
\xi_{yyy}=(-1)^r\sgm^3g\hat\vt_r,\eqno(6.78)$$
$$\eta_{yt}=\tau_t+(-1)^r(\sgm \zeta)_t
\vt_r+(-1)^r\sgm\sgm_t\zeta y\hat\vt_r,\;\;
\eta_{yyy}=\sgm^3\zeta\vt_r\eqno(6.79)$$ by (4.4) and (4.5).
 Thus
(4.2) and (4.3) are implied by the following system of partial
differential equations:
$$\sgm_x=\sgm_z=0,\qquad
\zeta\kappa_z-\sgm gf_x=0,\qquad g\tau_x-(-1)^r\sgm\zeta
f_x=0,\eqno(6.80)$$
$$\kappa_t+\kappa\kappa_x+\tau\kappa_z+(a\dlt_{0,r}+\dlt_{1,r})\sgm^2
gg_x+\frac{1}{\rho}p_x=0,\eqno(6.81)$$ $$\tau_t
+\kappa\tau_x+\tau\tau_z-(-1)^r(a\dlt_{0,r}+\dlt_{1,r})\sgm^2\zeta\zeta_z
+\frac{1}{\rho}p_z=0,\eqno(6.82)$$
$$(\sgm g)_t+(\sgm g\kappa)_x+(\sgm
g)_z\tau-(-1)^r\nu\sgm^3g=0,\eqno(6.83)$$ $$(\sgm \zeta)_t+(\sgm
\zeta\tau)_z+(\sgm
\zeta)_x\kappa-(-1)^r\nu\sgm^3\zeta=0,\eqno(6.84)$$
$$\sgm_t -\sgm(\kappa_x+\tau_z)=0,\eqno(6.85)$$
$$\zeta g_z-g\zeta_z=0,\qquad
g\zeta_x- g_x\zeta=0.\eqno(6.86)$$

By (6.80) and (6.86), we have
$$\sgm=e^\al, \qquad\zeta=g\tan\gm,\qquad
\tau_x\cot\gm-(-1)^r\kappa_z\tan\gm=0.\eqno(6.87)$$ Moreover,
(6.83) and (6.84) become
$$ g_t+\kappa g_x+\tau
g_z+(\al'+\kappa_x-(-1)^r\nu e^{2\al})g=0,\eqno(6.88)$$
$$\gm'g\sec^2\gm+(g_t+\kappa g_x+\tau g_z+(\al'+\tau_z
-(-1)^r\nu e^{2\al})g)\tan\gm=0.\eqno(6.89)$$ Furthermore,
$[(6.89)-\tan\gm\;\times (6.88)]/g$ yields
$$\gm'\sec^2\gm+(\tau_z-\kappa_x)\tan\gm=0\lra
\kappa_x-\tau_z=2\gm'\csc 2\gm.\eqno(6.90)$$ On the other hand,
(6.85) says
$$\kappa_x+\tau_z=\al'.\eqno(6.91)$$
Thus
$$\kappa_x=\frac{\al'}{2}+\gm'\csc 2\gm,\qquad\tau_z
=\frac{\al'}{2}-\gm'\csc 2\gm.\eqno(6.92)$$
 We take
$$\kappa=\left(\frac{\al'}{2}+\gm'\csc
2\gm\right)x,\qquad\tau=\left(\frac{\al'}{2}-\gm'\csc
2\gm\right)z\eqno(6.93)$$ for simplicity. So  the third equation
in (6.87) naturally holds. Moreover, the compatibility
$p_{xz}=p_{zx}$ in (6.81) and (6.82) is implied by $(g^2)_{xz}=0$.
So
$$g^2=\phi(t,x)+\psi(t,z)\eqno(6.94)$$
for some two-variable functions $\phi$ and $\psi$. By (6.93),
(6.88) becomes \begin{eqnarray*}\hspace{1cm}& &
(g^2)_t+\left(\frac{\al'}{2}+\gm'\csc 2\gm\right)x
(g^2)_x+\left(\frac{\al'}{2}-\gm'\csc 2\gm\right)z(g^2)_z\\
& &+(3\al'+2\gm'\csc 2\gm-(-1)^r2\nu
e^{2\al})g^2=0.\hspace{6.1cm}(6.95)\end{eqnarray*} Thus
$$g^2=e^{-3\al+(-1)^r2\nu\int e^{2\al}dt}(\iota(xe^{-\al/2}\sqrt{\cot\gm})+
\Im(ze^{-\al/2}\sqrt{\tan\gm}))\cot\gm\;\eqno(6.96)$$ for some
one-variable functions $\iota,\Im$. According (6.81), (6.82),
(6.87),  (6.93) and (6.96), we have:
\begin{eqnarray*}p&=&\frac{\rho}{2}\{(a\dlt_{0,r}+\dlt_{1,r})
[(-1)^r(\tan\gm)
\Im(ze^{-\al/2}\sqrt{\tan\gm})-(\cot\gm)\iota(xe^{-\al/2}\sqrt{\cot\gm})]
\\ & &\times e^{-\al+(-1)^r2\nu\int e^{2\al}dt}-\left[ \left(\frac{\al'}{2}+\gm'\csc
2\gm\right)^2+\frac{{\al'}'}{2}+\csc
2\gm\;({\gm'}'-2(\gm')^2\cot 2\gm)\right]x^2\\
& &- \left[ \left(\frac{\al'}{2}-\gm'\csc
2\gm\right)^2+\frac{{\al'}'}{2}-\csc 2\gm\;({\gm'}'-2(\gm')^2\cot
2\gm)\right]z^2\} \hspace{2.9cm}(6.97)
\end{eqnarray*}
modulo the transformation in (2.21). Furthermore, we take $f=0$ by
(6.80).

 Note $\xi=\kappa y+g\vt_r$ and $\eta=\tau y+\zeta\hat\vt_r$. By (4.1), we have the following theorem. \psp

{\bf Theorem 6.4}. {\it Let $\iota,\Im$ be any one-variable
functions and let $\al,\gm$ be any functions of $t$.  For $r=0,1$,
we define $\vt_r$ and $\hat\vt_r$ in (6.1) and (6.2) with
$\varpi=e^\al y$ and $a\in\mbb{R}$.
 We have the
following solutions of the three-dimensional classical non-steady
boundary layer equations (1.3), (1.4) and (1.5):
\begin{eqnarray*}\hspace{2cm}u&=&\left(\frac{\al'}{2}+\gm'\csc
2\gm\right)x+e^{-\al/2+(-1)^r\nu\int e^{2\al}dt}\hat\vt_r\\
& &\times\sqrt{(\iota(xe^{-\al/2}\sqrt{\cot\gm})+
\Im(ze^{-\al/2}\sqrt{\tan\gm}))\cot\gm},\hspace{3cm}(6.98)\end{eqnarray*}
\begin{eqnarray*}\hspace{2cm}w&=&\left(\frac{\al'}{2}-\gm'\csc
2\gm\right)z+(-1)^re^{-\al/2+(-1)^r\nu\int e^{2\al}dt}\vt_r\\
& &\times\sqrt{(\iota(xe^{-\al/2}\sqrt{\cot\gm})+
\Im(ze^{-\al/2}\sqrt{\tan\gm}))\tan\gm},\hspace{3cm}(6.99)\end{eqnarray*}
\begin{eqnarray*}\hspace{1cm}v&=&-\al'y-
\frac{e^{-2\al+(-1)^r\nu\int
e^{2\al}dt}}{2\sqrt{(\iota(xe^{-\al/2}\sqrt{\cot\gm})+
\Im(ze^{-\al/2}\sqrt{\tan\gm}))}}\\ & &\times[(\cot\gm)\vt_r
\iota'(xe^{-\al/2}\sqrt{\cot\gm})+(\tan\gm)\hat\vt_r
\Im'(ze^{-\al/2}\sqrt{\tan\gm})]
\hspace{2.1cm}(6.100)\end{eqnarray*} and $p$ is given in (6.97).}

\section{3-D Rational Approach}

In this section, we will find certain function-parameter exact
solutions of rational in $y$ for the three-dimensional classical
non-steady boundary layer equations (1.3), (1.4) and (1.5).

 First we consider another three cases of solving
the three-dimensional classical non-steady boundary layer
equations (1.3), (1.4) and (1.5) based on the settings in
(4.1)-(4.12). \psp

{\it Case 8}. $H=\Phi=y^{-1}$.\psp

In this case $\sgm=1$ and $\varpi=y$. So (4.11) becomes
\begin{eqnarray*}\hspace{1cm}& &\xi_y\xi_{yx}-(\xi_x+\eta_z)\xi_{yy}+\eta_y\xi_{yz}
=\kappa\kappa_x+\tau\kappa_z-[\zeta\kappa_z+(g\kappa)_x+g_z\tau]y^{-2}
\\ & &
-2g[(\kappa_x+\tau_z)y + f_x]y^{-3} + (\zeta
g_z-gg_x-2g\zeta_z)y^{-4}
 .\hspace{4.6cm}(7.1)
\end{eqnarray*}
Moreover,
$$\xi_{yt}=\tau_t-g_ty^{-2},\qquad \xi_{yyy}=-6 gy^{-4}.\eqno(7.2)$$
By symmetry, (4.2) and (4.3) are implied by the following system
of partial differential equations: $f_x=0$,
$$\kappa_t+\kappa\kappa_x+\tau\kappa_z+\frac{1}{\rho}p_x=0,\qquad
\tau_t+\kappa\tau_x+\tau\tau_z+\frac{1}{\rho}p_z=0,\eqno(7.3)$$
$$\zeta g_z-gg_x-2g\zeta_z+6\nu g=0,\qquad g\zeta_x-\zeta\zeta_z-2\zeta
g_x+6\nu\zeta=0,\eqno(7.4)$$
$$g_t+\zeta\kappa_z+\kappa
g_x+g_z\tau+g(3\kappa_x+2\tau_z)=0,\eqno(7.5)$$
$$\zeta_t+g\tau_x+\kappa
\zeta_x+\tau \zeta_z+\zeta(3\tau_z+2\kappa_x)=0.\eqno(7.6)$$

We take $f=0$ and assume $\zeta=g\tan\gm$ for some function $\gm$
of $t$. Use the notations in (4.31). Then both equations in (7.4)
are equivalent to the equation:
$$g_x+g_z\tan\gm=6\nu\lra g=6\nu x+\phi(t,\hat\varpi)\eqno(7.7)$$
for some two variable function $\phi$. Moreover,
$(7.6)-\tan\gm\times(7.5)$ is implied by
$$\gm'\sec^2\gm+\tau_x-\kappa_z\tan^2\gm+(\tau_z-\kappa_x)\tan\gm=0,
\eqno(7.8)$$ equivalently,
$$\gm'\sec\gm+\ptl_{\td\varpi}(\tau-\kappa\tan\gm)=0\eqno(7.9)$$
by (4.32). So
$$\tau=\kappa\tan\gm+\psi(t,\hat\varpi)-\gm'\td\varpi\sec\gm\eqno(7.10)$$
for another two-variable function $\psi$. To simplify the problem,
we assume
$$\kappa_z-\tau_x=0\sim
\ptl_{\hat\varpi}(\kappa)\sec\gm=-\psi_{\hat\varpi}\sin\gm-\gm'\eqno(7.11)$$
by (4.31) and (4.32). We take
$$\kappa=\al\td\varpi-\frac{1}{2}\psi\sin
2\gm-\gm'\hat\varpi\cos\gm\eqno(7.12)$$ for some function $\al$ of
$t$.

Note that (7.5) becomes
\begin{eqnarray*} \hspace{1cm}& &\phi_t-\gm'\td\varpi\phi_{\hat\varpi} +
6\nu\al\td\varpi-3\nu\psi\sin 2\gm-6\nu\gm'\hat\varpi\cos\gm
 +(\psi-\gm'\td\varpi\sec\gm)\phi_{\hat\varpi}\cos\gm\\ & &+
(6\nu
x+\phi)(3\al\sec\gm+2\psi_{\hat\varpi}\cos\gm-2\gm'\tan\gm)=0,
\hspace{4.3cm}(7.13)\end{eqnarray*}equivalently,
\begin{eqnarray*}
& &\phi_t-3\nu\psi\sin 2\gm+(\phi-6\nu\hat\varpi\sin\gm)
(3\al\sec\gm+2\psi_{\hat\varpi}\cos\gm-2\gm'\tan\gm)-6\nu\gm'\hat\varpi\cos\gm\\
&&+\phi_{\hat\varpi}\psi\cos\gm+2\td\varpi(12\nu\al-\gm'\phi_{\hat\varpi}+
6\nu\psi_{\hat\varpi}\cos^2\gm-6\nu\gm'\sin\gm)=0.\hspace{2.8cm}(7.14)\end{eqnarray*}
First we have:
$$12\nu\al-\gm'\phi_{\hat\varpi}+
6\nu\psi_{\hat\varpi}\cos^2\gm-6\nu\gm'\sin\gm=0.\eqno(7.15)$$ So
we take
$$\psi=\frac{1}{6\nu}\gm'\phi\sec^2\gm+\gm'\hat\varpi\tan\gm\;\sec\gm
-2\al\hat\varpi\sec^2\gm.\eqno(7.16)$$  Now (7.14) becomes
\begin{eqnarray*}
& &\phi_t+(\cos\gm\;\phi_{\hat\varpi}-3\nu\sin
2\gm)\left(\frac{1}{6\nu}\gm'\phi\sec^2\gm+\gm'\hat\varpi\tan\gm\;\sec\gm
-2\al\hat\varpi\sec^2\gm\right)\\ &
&-6\nu\gm'\hat\varpi\cos\gm+\frac{1}{3\nu}\gm'\phi_{\hat\varpi}
(\phi-6\nu\hat\varpi\sin\gm)\sec\gm=0,
\hspace{5.4cm}(7.17)\end{eqnarray*} equivalently.
\begin{eqnarray*}\hspace{1cm}&
&\phi_t+\frac{1}{2\nu}\gm'\phi\phi_{\hat\varpi}\sec\gm
-\gm'\phi\tan\gm-
(\gm'\tan\gm+2\al\sec\gm)\hat\varpi\phi_{\hat\varpi}
\\&&+6\nu(2\al\tan\gm-\gm'\sec\gm)\hat\varpi=0.\hspace{7.7cm}(7.18)\end{eqnarray*}
For simplicity, we take
$$\phi=\be\hat\varpi\eqno(7.19)$$
for a function $\be\neq 6\sin\gm$ of $t$. Then (7.18) gives
$$\al=\frac{\be^2\gm'+2\nu(\be'\cos\gm-2\be\gm'\sin\gm-
6\gm')}{4\nu(\be-6\nu\sin\gm)}.\eqno(7.20)$$ In summary, we have
$$g=6\nu x+\be\hat\varpi,\qquad
\zeta=(6\nu x+\be\hat\varpi)\tan\gm,\eqno(7.21)$$
\begin{eqnarray*}\hspace{1cm}\kappa&=&
\al\td\varpi-\left(\frac{1}{6\nu}\gm'\be\tan\gm+\gm'\sin\gm\;\tan\gm
-2\al\tan\gm\right)\hat\varpi-\gm'\hat\varpi\cos\gm\\
&=&\frac{\be^2\gm'+2\nu(\be'\cos\gm-2\be\gm'\sin\gm-
6\gm')}{4\nu(\be-6\nu\sin\gm)}
(\td\varpi+2\hat\varpi\tan\gm)\\
&
&-\left(\frac{1}{6\nu}\be\gm'\tan\gm+\gm'\sec\gm\right)\hat\varpi,
\hspace{7.2cm}(7.22)\end{eqnarray*}
\begin{eqnarray*}\hspace{1cm}\tau&=&\frac{\be^2\gm'+2\nu(\be'\cos\gm-2\be\gm'\sin\gm-
6\gm')}{4\nu(\be-6\nu\sin\gm)}
(\td\varpi\tan\gm-2\hat\varpi)\\
&
&+\frac{1}{6\nu}\be\gm'\hat\varpi-\gm'\td\varpi\sec\gm.\hspace{8.6cm}(7.23)\end{eqnarray*}
Set
\begin{eqnarray*}\hspace{1cm}\hat\kappa
&=&\frac{\be^2\gm'+2\nu(\be'\cos\gm-2\be\gm'\sin\gm-
6\gm')}{8\nu\cos\gm\;(\be-6\nu\sin\gm)}
(\td\varpi^2-2\hat\varpi^2)\\
& &+\frac{1}{12\nu}\be\gm'\sec\gm\;\hat\varpi^2+
\frac{\gm'}{2}\tan\gm(\hat\varpi^2-\td\varpi^2)-\gm'\hat\varpi\td\varpi.
\hspace{4cm}(7.24)\end{eqnarray*} Then
$$\kappa=\hat\kappa_x,\qquad \tau =\hat\kappa_z.\eqno(7.25)$$
By the fact $\kappa_z=\tau_x$, (7.11) and (7.24), we have
$$p=-\rho\hat\kappa_t-\frac{\rho}{2}(\kappa^2+\tau^2)\eqno(7.26)$$
modulo the transformation in (2.21).

 In this case, $\xi=\kappa
y+gy^{-1}$ and $\eta=\tau y+\zeta y^{-1}$.  By (4.1), we have the
following theorem. \psp

{\bf Theorem 7.1}. {\it Let $\be,\gm$ be any functions of $t$. In
terms of the notations in (4.31), we have the following solutions
of the three-dimensional classical non-steady boundary layer
equations (1.3), (1.4) and (1.5):
\begin{eqnarray*}\hspace{1cm}u&=&\frac{\be^2\gm'+2\nu(\be'\cos\gm-2\be\gm'\sin\gm-
6\gm')}{4\nu(\be-6\nu\sin\gm)}(\td\varpi+2\hat\varpi\tan\gm)\\
&
&-\left(\frac{1}{6\nu}\be\gm'\tan\gm+\gm'\sec\gm\right)\hat\varpi
-(6\nu x+\be\hat\varpi)y^{-2},\hspace{3.8cm}(7.27)\end{eqnarray*}
\begin{eqnarray*}\hspace{1cm}w&=&\frac{\be^2\gm'+2\nu(\be'\cos\gm-2\be\gm'\sin\gm-
6\gm')}{4\nu(\be-6\nu\sin\gm)}(\td\varpi\tan\gm-2\hat\varpi)\\
& &+\frac{1}{6\nu}\be\gm'\hat\varpi-\gm'\td\varpi\sec\gm- (6\nu
x+\be\hat\varpi)y^{-2}\tan\gm,\hspace{4.3cm}(7.28)\end{eqnarray*}
$$v=\frac{\be^2\gm'+2\nu(\be'\cos\gm-2\be\gm'\sin\gm-
6\gm')}{4\nu(\be-6\nu\sin\gm)\cos\gm}y-
\frac{\be\gm'}{6\nu}y\sec\gm-6\nu y^{-1} \eqno(7.29)$$ and $p$ is
given in (7.26) via (7.22)-(7.24).}\psp

{\it Case 9}. $H=\Phi=0$.\psp

In this case, $f$ can be any function of $t,x,z$, and (4.2) and
(4.3) are equivalent to the equations in (7.3), whose
compatibility $p_{xz}=p_{zx}$ is equivalent to (4.30). The
simplest solutions are
$$\kappa=\theta_x(t,x,z),\qquad \tau=\theta_z(t,x,z)\eqno(7.30)$$
for some three-variable function $\theta$. Since
$\kappa_z=\tau_x$,
$$p=-\rho\theta_t-\frac{\rho}{2}(\theta_x^2+\theta_z^2)\eqno(7.31)$$
modulo the transformation in (2.21). Besides, the $\kappa$ and
$\tau$ in Cases 1,3,4 and 8 work for this case.

 Next we assume
$$\kappa=e^{2\al} z+h_x(t,x,z),\qquad \tau=h_z(t,x,z),\eqno(7.32)$$
where $\al$ is a nonzero function of $t$ and $h$ is a function of
$t,x,z$. Now (4.30) becomes
$$2\al'+h_{xx}+h_{zz}=0.\eqno(7.33)$$
Hence
$$h=\Upsilon_x-\al' x^2\eqno(7.34)$$
for some time-dependant harmonic function $\Upsilon(t,x,z)$, that
is,
$$\Upsilon_{xx}+\Upsilon_{zz}=0.\eqno(7.35)$$
In this subcase,
$$\kappa=\Upsilon_{xx}-2\al'x+e^{2\al}
z,\qquad\tau=\Upsilon_{xz}.\eqno(7.36)$$
$$\kappa_t=\Upsilon_{xxt}-2{\al'}'x+2\al'e^{2\al}z, \eqno(7.37)$$
$$\kappa_x=\Upsilon_{xxx}-2\al',\qquad
\kappa_z=\Upsilon_{xxz}+e^{2\al}.\eqno(7.38)$$ Thus
\begin{eqnarray*} & &\kappa_t+\kappa\kappa_x+\tau \kappa_z =
\Upsilon_{xxt}-2{\al'}'x+2\al'e^{2\al}z\\
& &+(\Upsilon_{xx}-2\al'x+e^{2\al}
z)(\Upsilon_{xxx}-2\al')+\Upsilon_{xz}(\Upsilon_{xxz}+e^{2\al})
\\
&=&\Upsilon_{xtx}+\frac{1}{2}(\Upsilon_{xx}^2+\Upsilon_{xz}^2)_x-2\al'
(x\Upsilon_{xx})_x
+4(\al')^2x+e^{2\al}(z\Upsilon_{xxx}+\Upsilon_{zx})-2{\al'}'x
\hspace{0.6cm}(7.39)\end{eqnarray*} and
$$\tau_t+\kappa\tau_x+\tau\tau_z=\Upsilon_{xzt}+
(\Upsilon_{xx}-2\al' x+e^{2\al}
z)\Upsilon_{xxz}+\Upsilon_{xz}\Upsilon_{xzz}.\eqno(7.40)$$ So
$$p=-\rho\left[\Upsilon_{xt}+\frac{1}{2}(\Upsilon_{xx}^2+\Upsilon_{xz}^2)-2\al'
x\Upsilon_{xx}
+(2(\al')^2-{\al'}')x^2+e^{2\al}(z\Upsilon_{xx}+\Upsilon_z)\right]
\eqno(7.41)$$ modulo the transformation in (2.21).

Denote
$$\varpi=x^2+z^2.\eqno(7.42)$$
$$\kappa=\al'x+z\phi(t,\varpi),\qquad \tau=\al'z-x\phi(t,\varpi)\eqno(7.43)$$
for some functions $\al$ of $t$ and $\phi$ of $t$ and $\varpi$.
Note
$$\kappa_z-\tau_x=2(\varpi\phi)_\varpi.\eqno(7.44)$$
Then (4.30) becomes
$$(\varpi\phi)_{\varpi t}+2\al'[(\varpi\phi)_\varpi+
\varpi(\varpi\phi)_{\varpi\varpi}=0.\eqno(7.45)$$ Thus
$$\phi=\frac{\Im(e^{-2\al}\varpi)+\be}{\varpi}\eqno(7.46)$$
for a function $\be$ of $t$ and a one-variable function $\Im$. So
$$\kappa=\al'x+\frac{z(\Im(e^{-2\al}\varpi)+\be)}{\varpi},\qquad
\tau=\al'z-\frac{x(\Im(e^{-2\al}\varpi)+\be)}{\varpi}.\eqno(7.47)$$
Moreover, $$\kappa_t+\kappa\kappa_x+\tau\kappa_z
=({\al'}'+(\al')^2)x+\frac{\be'z}{\varpi}
-x\frac{(\Im(e^{-2\al}\varpi)+\be)^2}{\varpi^2},\eqno(7.48)$$
$$\tau_t+\kappa\tau_x+\tau\tau_z
=({\al'}'+(\al')^2)z-\frac{\be'x}{\varpi}
-z\frac{(\Im(e^{-2\al}\varpi)+\be)^2}{\varpi^2}.\eqno(7.49)$$
 Hence
\begin{eqnarray*}p&=&\frac{\rho}{2}\left\{e^{-2\al}\int
\frac{(\Im(e^{-2\al}(x^2+z^2))+\be)^2}{[e^{-2\al}(x^2+z^2)]^2}
\:d[e^{-2\al}(x^2+z^2)]-({\al'}'+(\al')^2)(x^2+z^2)\right\}
\\ & &+\rho\be'\arctan\frac{z}{x}\hspace{11.2cm}(7.50)\end{eqnarray*}
 modulo the transformation in (2.21).

The final subcase we are concerned with is to assume
$$\tau=\ves(t,z)\eqno(7.51)$$
for some two-variable function $\ves$. In this subcase, (4.30) is
equivalent to
$$\ptl_z(\kappa_t+\kappa\kappa_x+\tau\kappa_z)=0.\eqno(7.52)$$
We look for a solution of the form
$$\kappa=\phi(t,z)+\psi(t,z)x\eqno(7.53)$$
for some two-variable function $\phi$ and $\psi$. Note
$$\kappa_t+\kappa\kappa_x+\tau\kappa_z=
\phi_t+\psi_tx+\psi\phi+\psi^2x+\ves_z\phi_z+\ves_z\psi_zx.\eqno(7.54)$$
So (7.48) is equivalent to the following system of partial
differential equations:
$$\ptl_z(\phi_t+\psi\phi+\ves\phi_z)=0,\qquad
\ptl_z(\psi_t+\psi^2+\ves\psi_z)=0.\eqno(7.55)$$ To solve the
above system, we assume
$$\ves=\frac{\al}{\psi_z}-\frac{\vs_t(t,z)}{\vs_z(t,z)}\eqno(7.56)$$
for some functions $\al$ of $t$, and $\vs$ of $t$ and $z$. We have
the following solution of the above second equation:
$$\psi=\frac{1}{t+\Im(\vs)}\eqno(7.57)$$
for another one-variable function $\Im$. By the first equation,
$$\phi=\frac{b+\dlt_{\al,0}\Im_1(\vs)}{t+\Im(\vs)}\eqno(7.58)$$
for another one-variable function $\Im_1$ and a real constant $b$.
In this subcase,
$$\kappa=\frac{b+\dlt_{\al,0}\Im_1(\vs)+x}{t+\Im(\vs)},\qquad
\tau=-\frac{\al(t+\Im(\vs))^2}{\vs_z\Im'(\vs)}
-\frac{\vs_t}{\vs_z}.\eqno(7.59)$$ By (7.3) and (7.54),
\begin{eqnarray*}& &p=-\rho[\frac{1}{2}\left(\al x^2+
\left(\frac{\al(t+\Im(\vs))^2}{\vs_z\Im'(\vs)}
+\frac{\vs_t}{\vs_z}\right)^2\right)+\int
\left[\frac{\vs_t\vs_{zt}-\vs_{tt}\vs_z}{\vs_z^2}-
\frac{\al\vs_{zt}(t+\Im(\vs))^2}{\vs_z^2\Im'(\vs)}\right] dz
\\ & &
+\int\frac{(t+\Im(\vs))^2(\al\vs_t{\Im'}'(\vs)
-\al'\Im'(\vs))-2\al\Im'(\vs)(1+\vs_t\Im'(\vs))}
{\vs_z(\Im'(\vs))^2}dz+b\al x]\hspace{1.8cm}(7.60)\end{eqnarray*}
modulo the transformation in (2.21).

Recall $\xi=f+\kappa y$ and $\eta=\tau y$. By (4.1), we have the
following theorem. \psp

{\bf Theorem 7.2}.  {\it Let $\mu,\theta$ be  arbitrary function
of $t,x,z$ and let $\Im,\Im_1$ be any one-variable functions.
Suppose that $\al,\be$ are any functions of $t$, $\vs$ is any
function of $t$ and $z$, $b$ is a real constant and
$\Upsilon(t,x,z)$ is a time-dependent harmonic function in $x,z$.
Then we have the following solution of the three-dimensional
classical non-steady boundary layer equations (1.3), (1.4) and
(1.5): (1)
$$u=\theta_x,\qquad w=\theta_z,\qquad v=-(\theta_{xx}+\theta_{zz})y
+\mu\eqno(7.61)$$
and $p$ is given in (7.31); (2)
$$u=\Upsilon_{xx}-2\al'x+e^{2\al} z,\qquad w=\Upsilon_{xz},\qquad
v=\mu+2\al'y\eqno(7.62)$$ and $p$ is given in (7.41); (3)
$$u=\al'x+\frac{z(\Im(e^{-2\al}(x^2+z^2))+\be)}{x^2+z^2},
\qquad v=\mu-2\al'y, \eqno(7.63)$$
$$w=\al'z-\frac{x(\Im(e^{-2\al}(x^2+z^2))+\be)}{x^2+z^2},\eqno(7.64)$$
and $p$ is given (7.50); (4)
$$
u=\frac{(b+\dlt_{\al,0})\Im_1(\vs)+x}{t+\Im(\vs)}, \qquad
w=-\frac{\al(t+\Im(\vs))^2}{\vs_z\Im'(\vs)} -\frac{\vs_t}{\vs_z},
\eqno(7.65)$$
\begin{eqnarray*}\hspace{2cm}v&=&\frac{2\al\vs_z\Im'(\vs)(t+\Im(\vs))y}{\vs_z\Im'(\vs)}-
\frac{\al(t+\Im(\vs))^2(\vs_{zz}+\vs_z^2
{\Im'}'(\vs))y}{(\vs_z\Im'(\vs))^2}\\ &
&+\mu+\frac{\vs_{tz}\vs_z-\vs_t\vs_{zz}}{(\vs_z)^2}y-
\frac{y}{t+\Im(\vs)}\hspace{6.3cm}(7.66)\end{eqnarray*} and $p$ is
given in (7.60).}\psp

{\it Case 10}. $H=\Phi=y^2$. \psp

In this case, $\sgm=1$.  Now (4.11) becomes
\begin{eqnarray*}\hspace{1cm}& &\xi_y\xi_{yx}-(\xi_x+\eta_z)\xi_{yy}+\eta_y\xi_{yz}
 =\kappa_t+\kappa\kappa_x+\tau\kappa_z-2gf_x\\ & &+2[(g\kappa)_x+
g_z\tau+\zeta\kappa_z- g(\kappa_x+\tau_z)]y + 2(g
 g_x+2\zeta g_z-g\zeta_z)y^2.\hspace{2.3cm}(7.67)
\end{eqnarray*}
Moreover,
$$\xi_{yt}=\kappa_t+2g_ty,\qquad \xi_{yyy}=0.\eqno(7.68)$$
Modulo the transformation (2.22) and (2.23), we assume $\tau=0$.
By symmetry, (4.2) and (4.3) are equivalent to the following
system of partial differential equations,
$$\kappa_t+\kappa\kappa_x-2gf_x+\frac{1}{\rho}p_x=0,\qquad
-2\zeta f_x+\frac{1}{\rho}p_z=0,\eqno(7.69)$$
$$g_t+\kappa g_x
+\zeta\kappa_z=0,\qquad \zeta_t+
\zeta_x\kappa-\zeta\kappa_x=0,\eqno(7.70)$$
$$gg_x+2\zeta g_z-g\zeta_z=0,\qquad\zeta\zeta_z+2\zeta_x
 g-g_x\zeta=0.\eqno(7.71)$$
After some computations, we take
$$g=\al(t),\qquad \zeta=\be(t)\eqno(7.72)$$
for functions $\al,\be$ of $t$. So (7.71) naturally holds. By
(7.70), we take
$$\kappa=\frac{\be' x-\al' z}{\be}.\eqno(7.73)$$
The compatibility $p_{xz}=p_{zx}$ in (7.69) gives
$$2(\be\ptl_x-\al\ptl_z)(f_x)=
\frac{{\al'}'}{\be}.\eqno(7.74)$$ Thus
$$f_x=\frac{{\al'}'(\be x-\al z)}{2\be(\al^2+\be^2)}+\phi_{\hat\varpi}
(t,\hat\varpi),\qquad\hat\varpi=\al x+\be z\eqno(7.75)$$ for some
two-variable function $\phi$. Thus
$$p=2\rho
\phi(t,\hat\varpi)-\frac{\rho{\be'}'}{2\be}x^2+\frac{\rho{\al'}'(\al
x^2+2\be xz-\al z^2)}{2(\al^2+\be^2)}\eqno(7.76)$$ modulo the
transformation in (2.21).

Recall $\xi=f+\kappa y+g y^2$ and $\eta=\zeta y^2$.
 By (4.1), we have the following theorem.\psp

{\bf Theorem 7.3}. {\it Let $\al,\be$ be arbitrary function of $t$
and let $\phi(t,\hat\varpi)$ be an arbitrary two-variable
function. Then we have the following solution of the
three-dimensional classical non-steady boundary layer equations
(1.3), (1.4) and (1.5):
$$u=2\al y+\frac{\be' x-\al' z}{\be},\qquad w=2\be y,\qquad
v=\frac{{\al'}'(\al z-\be x)}{2\be(\al^2+\be^2)}-\phi_{\hat\varpi}
(t,\hat\varpi)-\frac{\be'y}{\be}\eqno(7.77)$$ and $p$ is given in
(7.76), where $\hat\varpi=\al x+\be z$.} \psp

By the transformation in (2.22) and (2.23),  we next consider the
solutions of (4.2) and (4.3) in the following form:
$$\xi=fy^3+gy+h,\qquad\eta=\kappa y^3
 +\tau y^2+\mu y,\eqno(7.78)$$
 where $f,g,h,\kappa,\tau,\mu$ are functions of $t,x,z$  with $f\neq
0$. Note
$$\xi_y=3fy^2+g,\qquad \xi_{yy}=6fy,\qquad
\xi_{yyy}=6f,\qquad\xi_{yt}=3f_ty^2+g_t,\eqno(7.79)$$
$$\xi_{yx}=3f_xy^2+g_x,\qquad
\xi_{yz}=3f_zy^2+g_z,\qquad\xi_x=f_xy^3+g_xy+h_x,\eqno(7.80)$$
$$\eta_y=3\kappa y^2+2\tau y+\mu,\qquad \eta_{yyy}=6\kappa,\qquad\eta_{yt}=3\kappa_ty^2
+2\tau_t y+\mu_t\eqno(7.81)$$
$$\eta_{yx}=3\kappa_xy^2+2\tau_x y+\mu_x,\;\;
\eta_{yz}=3\kappa_zy^2+2\tau_z y+\mu_z,\;\;\eta_z=\kappa_zy^3
+\tau_z y^2+\mu_z y.\eqno(7.82)$$ So we have
\begin{eqnarray*}& &\xi_y\xi_{yx}-(\xi_x+\eta_z)\xi_{yy}+\eta_y\xi_{yz}
\\&=&(3fy^2+g)(3f_xy^2+g_x)-6fy((f_x+\kappa_z)y^3+\tau_zy^2+(g_x+\mu_z)y+h_x)\\ & &+(3\kappa y^2+2\tau
y+\mu)(3f_zy^2+g_z)
\\&=&3(ff_x-2f\kappa_z+3f_z\kappa)y^4+6(\tau
f_z-f\tau_z)y^3+3(f_xg-fg_x-2f\mu_z+\kappa g_z+f_z\mu)y^2\\ &
&+(2\tau g_z-6fh_x)y+\mu
g_z+gg_x,\hspace{8.4cm}(7.83)\end{eqnarray*}
\begin{eqnarray*}& &\xi_y\eta_{yx}-(\xi_x+\eta_z)\eta_{yy}
+\eta_y\eta_{yz}\\
&=&(3fy^2+g)(3\kappa_xy^2+2\tau_xy+\mu_x)-((f_x+\kappa_z)y^3+\tau_zy^2+(g_x+\mu_z)y+h_x)\\
&&\times(6\kappa y+2\tau) +(3\kappa y^2+2\tau
y+\mu)(3\kappa_zy^2+2\tau_zy+\mu_z)
\\ &=&2(3f\tau_x+2\kappa_z\tau-f_x\tau)y^3+2(g\tau_x+\tau_z\mu-\tau
g_x-3\kappa h_x)y+g\mu_x+\mu\mu_z-2h_x\tau\\ &
&+3(3f\kappa_x-2f_x\kappa+\kappa\kappa_z)y^4+[3(f\mu_x+\kappa_xg+
\kappa_z\mu-2\kappa g_x-\kappa\mu_z)
+2\tau\tau_z]y^2.\hspace{0.7cm}(7.84)\end{eqnarray*} Thus the
equations (4.2) and (4.3) are implied by the following system of
partial differential equations:
$$ff_x-2f\kappa_z+3f_z\kappa=0,\qquad 3f\kappa_x-2f_x\kappa+\kappa\kappa_z
=0,\eqno(7.85)$$ $$ \tau f_z-f\tau_z=0, \qquad
3f\tau_x+2\kappa_z\tau-f_x\tau=0,\eqno(7.86)$$
$$f_t+f_xg-fg_x-2f\mu_z+\kappa g_z+f_z\mu=0,\eqno(7.87)$$
$$3(\kappa_t+f\mu_x+\kappa_xg+\kappa_z\mu-2\kappa g_x-\kappa\mu_z)
+2\tau\tau_z=0,\eqno(7.88)$$
$$\tau g_z-3fh_x=0,\qquad\tau_t+g\tau_x+\tau_z\mu-\tau g_x-3\kappa h_x=0,
\eqno(7.89)$$
$$g_t+\mu g_z+gg_x+\frac{1}{\rho}p_x=6\nu f,\qquad\mu_t+g\mu_x+\mu\mu_z-2h_x\tau+\frac{1}{\rho}p_z=6\nu \kappa.
\eqno(7.90)$$

Let $\gm$ be a function of $t$. Again we use the notations
$\hat\varpi=z\cos\gm-x\sin\gm$ and $\td\varpi=z\sin\gm+x\cos\gm$.
We take the following solutions of (7.85):
$$f=\phi(t,\hat\varpi)\cos\gm,\qquad
\kappa=\phi(t,\hat\varpi)\sin\gm\eqno(7.91)$$ for a two-variable
function $\phi$. Since $f_x+\kappa_z=0$, (7.86) is implied by
$$\tau=\al\phi(t,\hat\varpi)\eqno(7.92)$$
for a function $\al$ of $t$. Now (7.87) and (7.88) become
$$-\gm'\phi\sin\gm+[\phi_t+\phi_{\hat\varpi}(\mu\cos\gm-g\sin\gm)-\phi
g_x-2\phi\mu_z]\cos\gm+\phi g_z\sin\gm=0,\eqno(7.93)$$
$$3[(\gm'+\mu_x)\phi\cos\gm+(\phi_t+\phi_{\hat\varpi}(\mu\cos\gm-
g\sin\gm) -2 \phi g_x-\phi\mu_z)\sin\gm]
+2\tau\tau_z=0.\eqno(7.94)$$ Moreover
$\cos\gm\times(7.93)-3\sin\gm\times(7.94)$ yields
$$3[\gm'+\mu_x\cos^2\gm-g_z\sin^2\gm+(\mu_z-g_x)\sin\gm\;\cos\gm]
+2\al^2\phi_{\hat\varpi}\cos^2\gm=0,\eqno(7.95)$$ equivalently,
$$\ptl_{\td\varpi}(\mu\cos\gm-g\sin\gm)=
-\frac{2}{3}\al^2\phi_{\hat\varpi}\cos^2\gm-\gm'.\eqno(7.96)$$

According to (7.91) and (7.92), the first equation in (7.89) gives
$$h_x=\frac{\al}{3}g_z\sec\gm.\eqno(7.97)$$
Substituting it into the second equation in (7.89), we get
$$\al'\phi+\al[\phi_t+\phi_{\hat\varpi}(\mu\cos\gm-g\sin\gm)
-\phi\ptl_{\td\varpi}(g)\sec\gm]=0. \eqno(7.98)$$ Furthermore,
(7.93) can be written as
$$-\gm'\phi\sin\gm+
[\phi_t+\phi_{\hat\varpi}(\mu\cos\gm-g\sin\gm)]\cos\gm
-\phi\ptl_{\td\varpi}(g)-2\phi\ptl_z(\mu\cos\gm-g\sin\gm)=0.
\eqno(7.99)$$ The above two equations implies
$$\ptl_z(\cos\gm\;\mu-\sin\gm\;g)=
-\frac{1}{2}\left(\frac{\al'}{\al}\cos\gm+\gm'\sin\gm\right).\eqno(7.100)$$
So
$$\mu\cos\gm-g\sin\gm=
-\frac{z}{2}\left(\frac{\al'}{\al}\cos\gm+\gm'\sin\gm\right)+\ves(t,x)
\eqno(7.101)$$ for some two-variable function $\ves$. Substituting
it into (7.96), we obtain
$$-\frac{\sin\gm}{2}\left(\frac{\al'}{\al}\cos\gm+\gm'\sin\gm\right)+
\ves_x\cos\gm=-\frac{2}{3}\al^2\phi_{\hat\varpi}\cos^2\gm-\gm',\eqno(7.102)$$
which shows that both $\ves_x$ and $\phi_{\hat\varpi}$ are purely
a function of $t$. Thus we take
$$\phi=\be\hat\varpi\eqno(7.103)$$
for a function $\be$ of $t$. Hence we have
$$\mu\cos\gm-g\sin\gm=-\left(\frac{2}{3}\al^2\be\cos^2\gm
+\gm'\right)\td\varpi+\left(\frac{\al^2\be}{3}\sin
2\gm+\frac{\gm'}{2}\tan\gm-\frac{\al'}{2\al}\right)\hat\varpi
\eqno(7.104)$$ modulo the transformations in (2.17)-(2.20).

 In order to solve (7.90) and (7.98), we assume $\al=2$ and
$$\frac{2}{3}\al^2\be\cos^2\gm+2\gm'=0\lra
 \be=-\frac{3\gm'}{4\cos^2\gm}.\eqno(7.105)$$
Then
$$\mu\cos\gm-g\sin\gm=\gm'\td\varpi\lra\mu=g\tan\gm+
\gm'\td\varpi\sec\gm\eqno(7.106)$$ by (7.104). Moreover, (7.98)
becomes
$$\be'\hat\varpi
-\be\hat\varpi\ptl_{\td\varpi}(g)\sec\gm=0\lra\ptl_{\td\varpi}(g)=
\frac{\be'}{\be}\cos\gm=\frac{{\gm'}'}{\gm'}\cos\gm+2\gm'\sin\gm.
\eqno(7.107)$$ Thus
$$g=\left(\frac{{\gm'}'}{\gm'}\cos\gm+2\gm'\sin\gm\right)\td\varpi
+\psi(t,\hat\varpi)\eqno(7.108)$$ for some two-variable function
$\psi$. The compatibility $p_{xz}=p_{zx}$ in (7.90) becomes
$$(\mu_x-g_z)_t+(g(\mu_x-g_z))_x+(\mu(\mu_x-g_z))_z+2\gm'(\hat\varpi g_z)_x
\sec^3\gm=-6\nu\be\eqno(7.109)$$ by (7.97) and (7.105).

Observe that (7.106) and (7.108) yield
$$g_z=\frac{{\gm'}'}{\gm'}\sin\gm\;\cos\gm+2\gm'\sin^2\gm+
\psi_{\hat\varpi}\cos\gm,\qquad
\mu_x-g_z=\gm'-\psi_{\hat\varpi}\sec\gm.\eqno(7.110)$$ So
$$(\mu_x-g_z)_t={\gm'}'-\gm'\frac{\sin\gm}{\cos^2\gm}\psi_{\hat\varpi}
+\gm'\td\varpi\psi_{\hat\varpi\hat\varpi}\sec\gm- \psi_{\hat\varpi
t}\sec\gm,\eqno(7.111)$$
$$(g(\mu_x-g_z))_x+(\mu(\mu_x-g_z))_z=
\left(\frac{{\gm'}'}{\gm'}+3\gm'\tan\gm\right)
(\gm'-\psi_{\hat\varpi}\sec\gm)-\gm'
\td\varpi\psi_{\hat\varpi\hat\varpi}\sec\gm\eqno(7.112)$$ and
$$(\hat\varpi g_z)_x=-\left(\frac{{\gm'}'}{\gm'}\sin\gm\;\cos\gm+2\gm'\sin^2\gm+
(\psi_{\hat\varpi}+\hat\varpi\psi_{\hat\varpi\hat\varpi})\cos\gm\right)\sin\gm.
\eqno(7.113)$$ Thus (7.109) is equivalent to
\begin{eqnarray*}\hspace{1cm}& &-
\psi_{\hat\varpi
t}\sec\gm-2\gm'\hat\varpi\psi_{\hat\varpi\hat\varpi}\tan\gm\;\sec\gm
-\left(\frac{{\gm'}'}{\gm'}+6\gm'\tan\gm\right)
\psi_{\hat\varpi}\sec\gm\\ &
&+2{\gm'}'(1-\tan^2\gm)+(\gm')^2(3-4\tan^2\gm)\tan\gm=\frac{9\nu\gm'}{2\cos^2\gm},
\hspace{3.1cm}(7.114)\end{eqnarray*} which can be written as
\begin{eqnarray*}\hspace{0.5cm}& &
\psi_{\hat\varpi
t}+2\gm'\tan\gm\;\hat\varpi\psi_{\hat\varpi\hat\varpi}
+\left(\frac{{\gm'}'}{\gm'}+6\gm'\tan\gm\right) \psi_{\hat\varpi}\\
&= &2{\gm'}'(1-\tan^2\gm)\cos\gm+(\gm')^2(3-4\tan^2\gm)\sin\gm-
\frac{9\nu\gm'}{2\cos\gm} . \hspace{3cm}(7.115)\end{eqnarray*}
Hence
\begin{eqnarray*}\hspace{1cm}\psi_{\hat\varpi}&=&\frac{\cos^6\gm}{\gm'}
\Im'(\hat\varpi\cos^2\gm)+\frac{9\nu}{2}\ln(\sec\gm-\tan\gm)
\\ & &+
\int[2{\gm'}'(1-\tan^2\gm)\cos\gm+(\gm')^2(3-4\tan^2\gm)\sin\gm]dt.
\hspace{2.3cm}(7.116)\end{eqnarray*}
 for some one-variable
function $\Im$. Therefore,
\begin{eqnarray*}\hspace{1cm}\psi&=&\frac{\cos^4\gm}{\gm'}
\Im(\hat\varpi\cos^2\gm)+\frac{9\nu}{2}\hat\varpi\ln(\sec\gm-\tan\gm)
+\vf\\& &+\hat\varpi
\int[2{\gm'}'(1-\tan^2\gm)\cos\gm+(\gm')^2(3-4\tan^2\gm)\sin\gm]dt
\hspace{2.3cm}(7.117)\end{eqnarray*}
  for some function $\vf$ of
$t$. According to (7.108),
\begin{eqnarray*}g&=&\left(\frac{{\gm'}'}{\gm'}\cos\gm
+2\gm'\sin\gm\right)\td\varpi+\frac{\cos^4\gm}{\gm'}
\Im(\hat\varpi\cos^2\gm)+\frac{9\nu}{2}\hat\varpi\ln(\sec\gm-\tan\gm)
+\vf\\ & &+\hat\varpi
\int[2{\gm'}'(1-\tan^2\gm)\cos\gm+(\gm')^2(3-3\tan^2\gm)\sin\gm]dt,
\hspace{3.3cm}(7.118)\end{eqnarray*}
\begin{eqnarray*}\mu&=&[\frac{\cos^4\gm }{\gm'} \Im(\hat\varpi
\cos^2\gm)+\hat\varpi
\int[2{\gm'}'(1-\tan^2\gm)\cos\gm+(\gm')^2(3-4\tan^2\gm)\sin\gm]dt
\\ & &+\frac{9\nu}{2}\hat\varpi\ln(\sec\gm-\tan\gm)+\vf]\tan\gm
+\left(\frac{{\gm'}'}{\gm'}\sin\gm
+\gm'\frac{1+2\sin^2\gm}{\cos\gm}\right)\td\varpi,
\hspace{1.2cm}(7.119)\end{eqnarray*}
\begin{eqnarray*}h_x&=&3\nu\ln(\sec\gm-\tan\gm)+\frac{2}{3}
\{\int[2{\gm'}'(1-\tan^2\gm)\cos\gm +(\gm')^2(3-4\tan^2\gm)\sin\gm]dt \\
& &+ \left(\frac{{\gm'}'}{\gm'}
+2\gm'\tan\gm\right)\tan\gm+\frac{\cos^6\gm}
{\gm'}\Im'(\hat\varpi\cos^2\gm)\}.
\hspace{4.6cm}(7.120)\end{eqnarray*}

According to (7.91), (7.92), (7.97), (7.103), (7.105), (7.106) and
(7.108),
\begin{eqnarray*}& &g_t+\mu g_z+gg_x-6\nu f\\ &=&
\left({\gm'}'\sin\gm+\frac{{{\gm'}'}'\gm'-({\gm'}')^2+2(\gm')^4}
{(\gm')^2}\cos\gm\right)\td\varpi+({\gm'}'\cos\gm+2(\gm')^2\sin\gm)
\hat\varpi+\psi_t\\ &
&-\gm'\psi_{\hat\varpi}\td\varpi+\left(\frac{{\gm'}'}{\gm'}\cos\gm
+2\gm'\sin\gm\right)^2\td\varpi\sec\gm
+\left(\frac{{\gm'}'}{\gm'}\cos\gm+2\gm'\sin\gm\right)\psi\sec\gm\\
& & +({\gm'}'\cos\gm+2(\gm')^2\sin\gm)\td\varpi\tan\gm
+\gm'\psi_{\hat\varpi}\td\varpi+\frac{9\nu\gm'}{2\cos\gm}\hat\varpi\\
&=&\left(6({\gm'}'+(\gm')^2\tan\gm)\sin\gm+\frac{{{\gm'}'}'\gm'+2(\gm')^4}
{(\gm')^2}\cos\gm\right)\td\varpi+\left(\frac{{\gm'}'}{\gm'}
+2\gm'\tan\gm\right)\psi\\ & &+\psi_t
+({\gm'}'\cos\gm+2(\gm')^2\sin\gm)
\hat\varpi+\frac{9\nu\gm'}{2\cos\gm}\hat\varpi,\hspace{5.8cm}(7.121)
\end{eqnarray*}
\begin{eqnarray*}& &\mu_t+g\mu_x+\mu\mu_z-2h_x\tau-6\nu \kappa
\\
&=&2\gm'g\sec^2\gm+{\gm'}'\td\varpi\sec\gm+
2(\gm')^2\td\varpi\tan\gm\;\sec\gm+(\gm')^2\hat\varpi\sec\gm
\\ & &+
(g_t+gg_x+\mu g_z-6\nu f)\tan\gm+2\gm'g_z\sec^3\gm\\
&=&2[({\gm'}'\cos\gm+2(\gm')^2\sin\gm)\td\varpi
+\gm'\psi]\sec^2\gm+{\gm'}'\td\varpi\sec\gm+
2(\gm')^2\td\varpi\tan\gm\;\sec\gm\\
&&+(\gm')^2\hat\varpi\sec\gm+2[({\gm'}'\cos\gm+2(\gm')^2\sin\gm)\sin\gm
+\gm'\psi_{\hat\varpi}\cos\gm]\hat\varpi\sec^3\gm\\ &
&+(g_t+gg_x+\mu g_z-6\nu f)\tan\gm
\\ &=&3({\gm'}'+2(\gm')^2\tan\gm)\td\varpi\sec\gm
+((\gm')^2+2{\gm'}'\tan\gm+4(\gm')^2\tan^2\gm)\hat\varpi\sec\gm
\\ & &+2\gm'(\psi+\hat\varpi\psi_{\hat\varpi})\sec^2\gm
+(g_t+gg_x+\mu g_z-6\nu
f)\tan\gm.\hspace{4cm}(7.122)\end{eqnarray*} Expression (7.115)
says
\begin{eqnarray*}\hspace{1cm}& &
\left(\frac{{\gm'}'}{\gm'} +2\gm'\tan\gm\right)\psi+\psi_t
+({\gm'}'\cos\gm+2(\gm')^2\sin\gm)
\hat\varpi+\frac{9\nu\gm'}{2\cos\gm}\hat\varpi
\\ &=&\vf'+[{\gm'}'(3-2\tan^2\gm)\cos\gm+(\gm')^2\sin\gm\;(5-4\tan^2\gm)]
\hat\varpi\\ & & -2\gm'\hat\varpi\psi_{\hat\varpi}\tan\gm
-2\gm'\psi\tan\gm+\left(\frac{{\gm'}'}{\gm'}
+4\gm'\tan\gm\right)\vf.\hspace{3.3cm}(7.123)\end{eqnarray*} By
(7.121) and (7.123), we set
\begin{eqnarray*}\hspace{1cm}\hat g&=&
\frac{\td\varpi^2}{2}\left(6\sin\gm\;({\gm'}'+(\gm')^2\tan\gm)+\frac{{{\gm'}'}'\gm'+2(\gm')^4}
{(\gm')^2}\cos\gm\right)\sec\gm
\\ & &+\frac{\hat\varpi^2}{2}
[{\gm'}'(2\tan\gm-3\cot\gm)+(\gm')^2(4\tan^2\gm-5)]
\\ & &+2\gm'\hat\varpi\psi\sec\gm+\left[\vf'+\left(\frac{{\gm'}'}{\gm'}
+4\gm'\tan\gm\right)\vf\right]\td\varpi\sec\gm.\hspace{3.3cm}(7.124)
\end{eqnarray*}
Then
$$\hat g_x=g_t+\mu g_z+gg_x-6\nu f\eqno(7.125)$$
by (4.31). Moreover, (4.31) yields
$\ptl_z(F(t,\hat\varpi))=-\ptl_x(F(t,\hat\varpi))\cot\gm$ for any
function $F$ of $t$ and $\hat\varpi$. Furthermore, (7.122),
(7.124) and (7.126) imply
\begin{eqnarray*}\hspace{1cm}& &\mu_t+g\mu_x+\mu\mu_z-2h_x\tau-6\nu
\kappa-\hat g_z\\ &=& 3({\gm'}'+2(\gm')^2\tan\gm)\td\varpi\sec\gm
+((\gm')^2+2{\gm'}'\tan\gm\;+4(\gm')^2\tan^2\gm)\hat\varpi\sec\gm
\\ & &+2\gm'(\psi+\hat\varpi\psi_{\hat\varpi})\sec^2\gm
+(\tan\gm+\cot\gm)\{[{\gm'}'(3-2\tan^2\gm)\cos\gm\\
& &+(\gm')^2(5-4\tan^2\gm)\sin\gm] \hat\varpi
-2\gm'\hat\varpi\psi_{\hat\varpi}\tan\gm -2\gm'\psi\tan\gm\}\\
&=&3({\gm'}'+2(\gm')^2\tan\gm)\td\varpi\sec\gm
+((\gm')^2+2{\gm'}'\tan\gm+4(\gm')^2\tan^2\gm)\hat\varpi\sec\gm
\\ & &+[{\gm'}'(3-2\tan^2\gm)\csc\gm+(\gm')^2(5-4\tan^2\gm)\sec\gm]
 \hat\varpi
\\&=&6z(\csc2\gm+\sec^2\gm).
\hspace{9.2cm}(7.126)\end{eqnarray*}
 Therefore,
\begin{eqnarray*}p&=&-\rho\{3z^2(\csc2\gm+\sec^2\gm)
+\left[\vf'+\left(\frac{{\gm'}'}{\gm'}
+4\gm'\tan\gm\right)\vf\right]\td\varpi\sec\gm\\ &
&+2\gm'\hat\varpi\psi\sec\gm +\frac{\hat\varpi^2}{2}
[{\gm'}'(2\tan\gm-3\cot\gm)+(\gm')^2(4\tan^2\gm-5)]
\\ & &+
\frac{\td\varpi^2}{2}\left(6({\gm'}'+(\gm')^2\tan\gm)\sin\gm+\frac{{{\gm'}'}'\gm'+2(\gm')^4}
{(\gm')^2}\cos\gm\right)\sec\gm \}\hspace{2.6cm}(7.127)
\end{eqnarray*}
modulo the transformation in (2.21).

By (4.1) and (7.78), we have the following theorem.\psp

{\bf Theorem 7.4}. {\it Let $\gm,\vf$ be arbitrary functions of
$t$ and let $\Im$ be an arbitrary one-variable function. Denote
$\hat\varpi=z\cos\gm-x\sin\gm$ and $\td\varpi=z\sin\gm+x\cos\gm$.
We have the following solutions of the three-dimensional classical
non-steady boundary layer equations (1.3), (1.4) and (1.5):
\begin{eqnarray*}u&=&\left(\frac{{\gm'}'}{\gm'}\cos\gm
+2\gm'\sin\gm\right)\td\varpi+\frac{\cos^4\gm}{\gm'}
\Im(\hat\varpi\cos^2\gm)+\frac{9\nu}{2}\hat\varpi\ln(\sec\gm-\tan\gm)
+\vf\\ & &+\hat\varpi
\int[2{\gm'}'(1-\tan^2\gm)\cos\gm+(\gm')^2(3-3\tan^2\gm)\sin\gm]dt
-\frac{9\gm'\hat\varpi}{4\cos\gm}y^2,\hspace{1cm}(7.128)\end{eqnarray*}
\begin{eqnarray*}w&=&[\frac{\cos^4\gm }{\gm'} \Im(\hat\varpi
\cos^2\gm)+\hat\varpi
\int[2{\gm'}'(1-\tan^2\gm)\cos\gm+(\gm')^2(3-4\tan^2\gm)\sin\gm]dt
\\ & &+\frac{9\nu}{2}\hat\varpi\ln(\sec\gm-\tan\gm)+\vf]\tan\gm
+\left(\frac{{\gm'}'}{\gm'}\sin\gm
+\gm'\frac{1+2\sin^2\gm}{\cos\gm}\right)\td\varpi\\ & & -
3\gm'\hat\varpi y\sec^2\gm
-\frac{9\gm'\hat\varpi\sin\gm}{4\cos^2\gm}y^2,
\hspace{7.9cm}(7.129)\end{eqnarray*}
\begin{eqnarray*}\hspace{1cm}v&=&\frac{3}{2}\gm'y^2\sec\gm
-\left(\frac{{\gm'}'}{\gm'}+3\gm'\tan\gm\right)y
-3\nu\ln(\sec\gm-\tan\gm)\\ & &-\frac{2}{3}
\{\int[2{\gm'}'(1-\tan^2\gm)\cos\gm +(\gm')^2(3-4\tan^2\gm)\sin\gm]dt \\
& &+ \left(\frac{{\gm'}'}{\gm'}
+2\gm'\tan\gm\right)\tan\gm+\frac{\cos^6\gm}
{\gm'}\Im'(\hat\varpi\cos\gm)\}.
\hspace{4cm}(7.130)\end{eqnarray*} and $p$ is given (7.127) via
(7.117).}\psp

Finally, we consider
$$\xi=f y^3+g y^2+\kappa y+h,\qquad \eta=\tau y+\zeta
y^{-1},\eqno(7.131)$$ where $f,g,h,\kappa,\tau$ and $\zeta$ are
functions of $t,x,z$ with $\zeta\neq 0$. First,
$$\xi_y=3fy^2+2gy+\kappa,\qquad
\xi_{yy}=6fy+2g,\qquad\xi_{yyy}=6f,\eqno(7.132)$$
$$\xi_{yt}=3f_ty^2+2g_ty+\kappa_t,\qquad\xi_{yx}=3f_xy^2+2g_xy+\kappa_x,
\eqno(7.133)$$ $$\xi_{yz}=3f_zy^2+2g_zy+\kappa_z,\qquad \xi_x=f_x
y^3+g_x y^2+\kappa_x y+h_x,\eqno(7.134)$$
$$\eta_y=\tau-\zeta y^{-2},\qquad \eta_{yy}=2\zeta
y^{-3},\qquad\eta_{yyy}=-6\zeta
y^{-4},\qquad\eta_{yx}=\tau_x-\zeta_xy^{-2}\eqno(7.135)$$
$$\eta_{yt}=\tau_t-\zeta_ty^{-2},\qquad\eta_{yz}=\tau_z-\zeta_zy^{-2},
\qquad\eta_z=\tau_z y+\zeta_z y^{-1}.\eqno(7.136)$$ So we have
\begin{eqnarray*}& &\xi_y\xi_{yx}-(\xi_x+\eta_z)\xi_{yy}+\eta_y\xi_{yz}
=(3fy^2+2gy+\kappa)(3f_xy^2+2g_xy+\kappa_x)\\ & &-(6fy+2g) (f_x
y^3+g_x y^2+(\kappa_x+\tau_z)y+h_x+\zeta_zy^{-1}) +(\tau-\zeta
y^{-2})(3f_zy^2+2g_zy+\kappa_z)\\
&&=3ff_xy^4+4f_xgy^3+(3f_x\kappa+2gg_x-3f\kappa_x
-6f\tau_z+3f_z\tau)y^2-2(g\zeta_z+g_z\zeta)y^{-1}\\ &
&+2(g_x\kappa-3fh_x-g\tau_z+g_z\tau)y+\kappa\kappa_x-6f\zeta_z-2gh_x
+\tau\kappa_z-3f_z\zeta-\zeta\kappa_zy^{-2}, \hspace{0.6cm}(7.137)
\end{eqnarray*}
\begin{eqnarray*}& &\xi_y\eta_{yx}-(\xi_x+\eta_z)\eta_{yy}
+\eta_y\eta_{yz} =(3fy^2+2gy+\kappa)(\tau_x-\zeta_xy^{-2})\\
& &-2\zeta y^{-3}(f_x y^3+g_x
y^2+(\kappa_x+\tau_z)y+h_x+\zeta_zy^{-1})+(\tau-\zeta
y^{-2})(\tau_z-\zeta_zy^{-2})\\
&&=3f\tau_xy^2+2g\tau_xy+\kappa\tau_x+\tau\tau_z-3f\zeta_x-2f_x\zeta
-2(g\zeta)_xy^{-1}\\ & &-(\kappa\zeta_x+2\zeta\kappa_x
+3\zeta\tau_z+\tau\zeta_z)y^{-2}-2h_x\zeta
y^{-3}-\zeta\zeta_zy^{-4}.\hspace{4.7cm}(7.138)
\end{eqnarray*}
Hence (4.2) and (4.3) are implied by the following system of
partial differential equations:
$$f_x=(g\zeta)_z=\kappa_z=\tau_x=(g\zeta)_x=h_x=0,\qquad\zeta_z=6\nu,
\eqno(7.139)$$ $$ f_t+\frac{2}{3}gg_x-f\kappa_x
-2f\tau_z+f_z\tau=0,\eqno(7.140)$$
$$g_t+g_x\kappa-g\tau_z+g_z\tau=0,\eqno(7.141)$$
$$\kappa_t+\kappa\kappa_x-6f\zeta_z-3f_z\zeta+\frac{1}{\rho}p_x=6\nu f,
 \eqno(7.142)$$
$$\tau_t+\tau\tau_z-3f\zeta_x+
\frac{1}{\rho}p_z=0,\eqno(7.143)$$
$$\zeta_t+\kappa\zeta_x+2\zeta\kappa_x
+3\zeta\tau_z+\tau\zeta_z=0.\eqno(7.144)$$

By (7.139), we take
$$h=0,\qquad f=\phi(t,z),\qquad\zeta=6\nu z+\psi(t,x),\qquad
g=\al\zeta^{-1}\eqno(7.145)$$ for some two-variable functions
$\phi,\psi$ and a function $\al$ of $t$. Moreover, (7.144) becomes
$$\psi_t+\kappa\psi_x+2\psi\kappa_x+12\nu
z\kappa_x+3\psi\tau_z+18\nu z\tau_z+6\nu\tau=0.\eqno(7.146)$$
Modulo the transformation in (2.19) and (2.20), we assume $\psi=0$
or $\psi_x\neq 0$. Note the compatibility $p_{xz}=p_{zx}$ in
(7.142) and (7.143) gives
$$f\zeta_{xx}=14\nu f_z+\zeta f_{zz}\lra f\psi_{xx}-\psi
f_{zz}=20\nu f_z+6\nu z f_{zz}.\eqno(7.147)$$ Furthermore, the
last equation in (7.145) says that
$g\times(7.144)+\zeta\times(7.141)$ yields
$$\al'+2\al(\kappa_x+\tau_z)=0\eqno(7.148)$$

{\it Subcase 1}. $\psi_x\neq 0$.\psp

In this subcase, applying $\ptl_x\ptl_z$ to (7.146) gives
$$4\nu\kappa_{xx}+\psi_x\tau_{zz}=0.\eqno(7.149)$$
So $\tau$ is a quadratic polynomial in $z$. The coefficients of
$z^2$ in (7.146) say that $\tau_{zz}=0$, and so $\kappa_{xx}=0$.
By the coefficients of $z$ in (7.146), we have
$$\kappa=-2\be' x,\qquad \tau=\be'z\eqno(7.150)$$
for some function $\be$ of $t$ modulo the transformations in
(2.17)-(2.20). Now (7.146) becomes
$$\psi_t-2\be 'x\psi_x-\be'\psi=0.
\eqno(7.151)$$ Hence
$$\psi=e^{\be}\Im(e^{2\be}x)\eqno(7.152)$$
for some one-variable function $\Im$.
 According to (7.147), we
take $f=0$. Furthermore, (7.140) gives us $g=0$. \pse

{\it Subcase 2}. $\psi= 0$.\psp

In this subcase, (7.146) becomes
$$2z\kappa_x+3z\tau_z+\tau=0.\eqno(7.153)$$
So
$$\kappa_x=-2\be',\qquad\tau=\gm z^{-1/3}+\be' z\eqno(7.154)$$
for some functions $\be$ and $\gm$. Moreover, (7.148) yields
$$\al=a\dlt_{\gm,0}e^{2\be}\qquad a\in\mbb{R}.\eqno(7.155)$$
Now (7.140) becomes
$$ f_t+\frac{2\gm}{3}z^{-4/3}f+(\gm z^{-1/3}+\be' z)f_z=0\eqno(7.156)$$
and (7.147) gives
$$10f_z+3 z f_{zz}=0.\eqno(7.157)$$
These two equations force us to take $\gm=0$ and
$$f=be^{7\be/3}z^{-7z/3}+c,\qquad b,c\in\mbb{R}.\eqno(7.158)$$

By (4.1) and (7.131), we have the following theorem:\psp

{\bf Theorem 7.5}. {\it Let $\be$ be arbitrary function of $t$ and
let $a,b,c$ be any real constants. Suppose that $\Im$ is any
one-variable function. We have the following solutions of the
three-dimensional classical non-steady boundary layer equations
(1.3), (1.4) and (1.5): (1)
$$u=-2\be' x,\;\;v=\be'y-6\nu y^{-1},\;\;w=\be'z-[6\nu z+e^{\be}\Im
(e^{2\be}x)]y^{-2}, \eqno(7.159)$$
$$p=\rho({\be'}'-2(\be')^2)x^2
-\frac{\rho}{2}({\be'}'+(\be')^2)z^2.\eqno(7.160)$$

 (2)
$$u=3(be^{7\be/3}z^{-7z/3}+c)y^2+\frac{ae^{2\be}y}{3\nu
z}-2\be'x,\;\;v=\be'y-6\nu y^{-1},\;\;w=(\be' -6\nu
y^{-2})z,\eqno(7.161)$$
$$p=\rho[42\nu
cx+({\be'}'-2(\be')^2)x^2]-\frac{\rho}{2}({\be'}'+(\be')^2)z^2.
\eqno(7.162)$$}

 \vspace{0.2cm}

\noindent{\Large \bf References}

\begin{description}

\item[{[DD]}] M. A. Daragan and  S. A. Derbenev, Group
investigations of laminar boundary layer on rotating wing, {\it
Izvestiya Vuzov. Aviatsionaya Tekhnika} {\bf 4} (1985), 81.

\item[{[D]}] S. A. Derbenev, Group properties of equations
boundary layer with magnetic field and chemical reactions, {\it
Trudy Kazan. Aviatsion. Inst.} {\bf 119} (1970), 100.

\item[{[G]}] K. G. Garaev, Group properties of an equation for
nonstationary space boundary layer of incompressible fluid, {\it
Trudy Kazan. Aviatsion. Inst.} {\bf 119} (1970), 47.

\item[{[GP]}] K. G. Garaev and Y. G. Pavlov, Group properties of
equations of optimally controlled boundary layer, {\it Izvestiya
Vuzov. Aviatsion. Tekhnika} {\bf 4} (1970), 5.

\item[{[I]}] N. H. Ibragimov, {\it Lie Group Analysis of
Differential Equations}, Volume 2, CRC Handbook, CRC Press, 1995.

\item[{[N]}] Nguen Van D'ep, On boundary layer equations for
fluids with a moment stress, {\it Prikladn. Matemanka i Mekhnika}
{\bf 32} (1989), no. 4, 748.

\item[{[Lm]}] M. Ya Lankerovich, Group properties of
three-dimensional boundary layer equations on arbitrary surfaces,
{\it Dinamika Splash. Sredi. Institute of Hydrodynamics} {\bf 7}
(1971), 12.

\item[{[LRT]}] C. C. Lin, E. Reissner and H. S. Tsien, On
two-dimensional non-steady motion of a slender body in a
compressible fluid, {\it J. Math. Phys.} {\bf 27} (1948), no. 3,
220.

\item[{[Ll]}] L. G. Loitsyanskii, {\it Laminar Boundary Layer},
Fizmatgiz, Moscow, 1962.

\item[{[M1]}] E. V. Mamontov, On the theory of nonstationary
transonic flows, {\it Dokl. Acad. Nauk SSSR} {\bf 185} (1969), no.
3, 538.

\item[{[O]}] L. V. Ovsiannikov, {\it Group Analysis of
Differential Equations}, Academic Press, New York, 1982.

\item[{[V]}] L. I. Vereshchagina, Group stratification of space
nonstationary boundary layer equations, {\it Vestnik Leningr.
Universiteta} {\bf 13} (1973), no. 3, 82.

\item[{[X]}] X. Xu,  Stable-Range approach to the equation of
nonstationary transonic gas flows, {\it Quart. Appl. Math}, to
appear.

\end{description}

\end{document}